# Federated Byzantine Quorum Systems
# (Extended Version)

## Álvaro García-Pérez
IMDEA Software Institute, Madrid, Spain

## Alexey Gotsman
IMDEA Software Institute, Madrid, Spain

—— **Abstract** ——————————————————————————

Some of the recent blockchain proposals, such as Stellar and Ripple, use quorum-like structures typical for Byzantine consensus while allowing for open membership. This is achieved by constructing quorums in a decentralised way: each participant independently chooses whom to trust, and quorums arise from these individual decisions. Unfortunately, the theoretical foundations underlying such blockchains have not been thoroughly investigated. To close this gap, in this paper we study decentralised quorum construction by means of federated Byzantine quorum systems, used by Stellar. We rigorously prove the correctness of basic broadcast abstractions over federated quorum systems and establish their relationship to the classical Byzantine quorum systems. In particular, we prove correctness in the realistic setting where Byzantine nodes may lie about their trust choices. We show that this setting leads to a novel variant of Byzantine quorum systems where different nodes may have different understanding of what constitutes a quorum.

## 1 Introduction

Blockchains are distributed databases that maintain an append-only ledger over a set of potentially Byzantine nodes. The nodes use a Byzantine fault-tolerant (BFT) consensus protocol to agree on a total order in which transactions are appended to the ledger. Blockchains usually come in two flavours. *Permissioned* blockchains assume a known set of participants, and are often based on classical BFT consensus protocols, such as PBFT [5]. In these protocols consensus is reached once a *quorum* of participants agrees on the same decision. Quorums can be defined as sets containing enough nodes in the system, e.g., $2f + 1$ out of $3f + 1$, assuming at most $f$ failures. They can also be defined by a more general structure called a *Byzantine quorum system (BQS)* [9]—a pair of a *quorum system* $\mathcal{Q}$ and a *fail-prone system* $\mathcal{B}$, the latter containing the sets of nodes characterising possible failure scenarios [9]. A subclass of *dissemination quorum systems (DQS)* requires $\mathcal{Q}$ and $\mathcal{B}$ to satisfy certain additional axioms, e.g., that any two quorums must intersect in a non-faulty node.

The other kind of blockchains are *permissionless* ones, which allow anyone to participate in consensus, with the membership constantly changing. These blockchains are often based on consensus protocols such as proof-of-work [7, 11], which do not rely on quorums. Despite the advantages of open membership, proof-of-work suffers from a number of drawbacks, including high energy consumption and the absence of hard guarantees on when a transaction can be considered successfully appended to the ledger. This has motivated a number of proposals of alternative architectures for permissionless blockchains. In this paper we focus on an intriguing new class of blockchains, such as Stellar [10] and Ripple [12], that use quorum-like structures typical for BFT consensus while allowing open membership.

This is achieved by constructing the quorum system in a decentralised way: each protocol participant independently chooses whom to trust, and quorums arise from these individual decisions.

In particular, in Stellar trust assumptions are specified using a *federated Byzantine quorum system (FBQS)*[1], where each participant selects a set of *quorum slices*—sets of nodes each of which would convince the participant to accept the validity of a given statement. Quorums are defined as sets of nodes $U$ such that each node in $U$ has some quorum slice fully within $U$, so that the nodes in a quorum can potentially reach an agreement. Consensus is then implemented by a fairly intricate protocol whose key component is *federated voting*—a protocol similar to Bracha's protocol for reliable Byzantine broadcast [2, 3]. Importantly, the decentralised nature of Stellar means that its participants operate based on an incomplete and inconsistent information about the trust assumptions of various nodes. First, a node may not have complete knowledge of quorum slices of all other nodes in the system. Second, Byzantine nodes may lie to other nodes about their choices of quorum slices. These features make establishing the correctness of the protocols underlying Stellar nontrivial.

Even though Stellar has been deployed as a functioning blockchain, its theoretical foundations remain shaky. The core Stellar protocols lack rigorous formalisations and proofs of correctness. Furthermore, as observed in [4], the federated quorum structures arising in Stellar are similar to the classical Byzantine quorum systems, where trust assumptions are defined globally for the whole system [9]. But the relationship between these has not been investigated, and this is a barrier to transferring ideas between the classical and the federated settings. In this paper we take the first step towards closing these gaps and perform a rigorous theoretical study of the basic concepts underlying the Stellar blockchain.

In more detail, the closest protocol to Stellar's core federated voting protocol in the world of classical Byzantine quorum systems is *Bracha broadcast* [2]. This implements *reliable Byzantine broadcast* [3] in a system of $3f + 1$ processes where at most $f$ processes can fail and any $2f + 1$ processes constitute a quorum. The protocol allows a node to send a message to a group of receivers; the sender may be faulty and can thus send different messages to different receivers. The protocol guarantees, among other properties, that: *(i)* correct receivers cannot deliver different messages; and *(ii)* if a correct receiver delivers a message, then eventually all correct receivers will deliver the same message [3]. Bracha's protocol can also be executed over an arbitrary dissemination quorum system (which we prove in §2).

We investigate Stellar's federated voting through the prism of the reliable Byzantine broadcast abstraction. We first consider an idealised setting (used in the Stellar whitepaper [10]) where nodes have consistent information about the quorum slices of nodes they interact with, i.e., faulty nodes do not equivocate about their quorum slices, but can otherwise behave arbitrarily. Thus, all nodes act according to the same FBQS. Our first contribution is to show that, in this setting, a federated Byzantine quorum system used in Stellar induces a classical Byzantine quorum system $(\mathcal{Q}, \mathcal{B})$, so that protocols such as Bracha broadcast can also be run out of the box in a federated setting (§3). However, Bracha broadcast requires a node to know the fail-prone system $\mathcal{B}$, and computing it from an FBQS would require the node to know the quorum slices of *all* the nodes in the system. This is infeasible in a system with open membership. Stellar circumvents this problem by allowing a node to take decisions based solely on the quorum slices of the nodes it interacts with directly. Our next contribution is to prove that the resulting *Stellar broadcast* implements what we call *weakly reliable Byzantine broadcast* (§4). This guarantees the safety property (i) of the usual reliable Byzantine broadcast and the liveness property (ii) where only certain *intact nodes* are guaranteed to deliver the message. In a given protocol execution, the set of intact nodes is the largest set that excludes all faulty nodes and satisfies certain basic properties allowing correct protocol operation among these nodes, e.g., that every two quorums intersect. This result strengthens the correctness theorems in the Stellar whitepaper [10], which only guarantee safety to intact nodes (instead of all correct ones) and guarantee liveness only under an assumption that an intact node delivers a message (rather than only correct one). We also show that a variation of the Stellar protocol actually implements the usual reliable Byzantine broadcast and prove that this variation is observationally equivalent to Bracha broadcast over the corresponding DQS. On a conceptual level, our correctness results show that some of the ad hoc protocols proposed for Stellar actually implement a variant of a well-known broadcast abstraction.

---

[1] In the Stellar whitepaper these are called *federated Byzantine agreement systems* [10]. The name used in this paper is chosen to be in line with common terminology [9] and to emphasise that these systems can be used for purposes other than solving consensus.

We next consider a more realistic setting where faulty nodes may lie about their choices of quorum slices (§5), which has not been covered by the existing correctness theorems for Stellar [10]. In this setting, different nodes may have different understanding of what the quorum slices of a given faulty node are. We capture this by the notion of a *subjective FBQS*, where each node has its own, possibly different, copy of an FBQS such that the different FBQSes agree on the quorum slices of correct nodes. This also leads different nodes to disagree on what constitutes a quorum. More precisely, a subjective FBQS induces a *subjective quorum system*, where each node has its own copy of a quorum system such that the projections of different quorum systems to correct nodes coincide. Our next result is to show that the Stellar broadcast correctly implements weakly reliable Byzantine broadcast even when each node acts according to its subjective view on the trust choices in the system. This result is required to get confidence in the correctness of the Stellar blockchain, since in its intended deployment faulty nodes may lie about everything, including their trust choices.

Finally, we generalise the theory of classical Byzantine quorum systems to allow for nodes to have inconsistent information about quorums, regardless of the origin of this inconsistency (§6); this may be useful in the future for dealing with approaches to decentralising trust different from Stellar. To this end, we introduce the notion of a *subjective dissemination quorum system*, where each node has its own Byzantine quorum system such that the different systems satisfy a variation on the DQS axioms. For example, we require that any two quorums, *even when coming from the systems of different nodes*, have to intersect in a non-faulty node. We show that Bracha broadcast over a subjective DQS implements reliable Byzantine broadcast. Furthermore, we show that a subjective FBQS induces a subjective DQS, on top of which one can execute Bracha broadcast.

Our results open the door to a deeper theoretical study of the novel quorum structures arising in a federated setting and to proving the correctness of the whole Stellar consensus protocol. We also hope that the connections we have established between classical broadcast abstractions and the novel Stellar protocols will enable transferring ideas from the existing optimised protocols for BFT consensus [6, 8] into the federated setting.

Proofs of the theorems stated in the paper are given in appendices.

## 2     Byzantine Quorum Systems and Bracha Broadcast

**System model.**   In this paper we consider a system consisting of a set of *client nodes* **C** and a disjoint set of *server nodes* **V**. We assume a Byzantine failure model: some nodes may be *faulty*, in which case they can deviate arbitrarily from their specification. All other nodes are called *correct*. We assume that any two nodes (clients or servers) can communicate over an asynchronous point-to-point channel. If both endpoints of the channel are correct, then this channel is both authenticated and reliable: a correct node receives a message from another correct node if and only if the latter node sent it.

**Byzantine quorum systems.**   A *quorum system* is a non-empty set $\mathcal{Q}$ of subsets of some universe of nodes, such that

- every quorum in $\mathcal{Q}$ is non-empty;
- any two quorums in $\mathcal{Q}$ intersect: $\forall U_1, U_2 \in \mathcal{Q}.\, U_1 \cap U_2 \neq \emptyset$;
- $\mathcal{Q}$ is closed under union: $\forall U_1, U_2 \in \mathcal{Q}.\, U_1 \cup U_2 \in \mathcal{Q}$.

In this paper we let the universe of nodes coincide with the set of servers **V**, so that $\mathcal{Q} \subseteq 2^{\mathbf{V}}$. A *fail-prone system* $\mathcal{B}$ is a non-empty set of subsets of **V** such that none of its elements is contained in another. A fail-prone system characterises the failure scenarios that can occur: in any execution of the system, some $B \in \mathcal{B}$ is meant to contain all the faulty servers. A *Byzantine quorum system (BQS)* is a pair $(\mathcal{Q}, \mathcal{B})$ of a quorum system and a fail-prone system. We sometimes use the adjective 'classical' to distinguish such quorum systems from federated ones that we define in §3.

In this paper we focus on a particular class of BQSes that satisfy additional properties. Namely, a *dissemination quorum system (DQS)* is a BQS with the following properties [9]:

- *D-consistency:* $\forall U_1, U_2 \in \mathcal{Q}.\, \forall B \in \mathcal{B}.\, U_1 \cap U_2 \nsubseteq B$; and
- *D-availability:* $\forall B \in \mathcal{B}.\, \exists U \in \mathcal{Q}.\, U \cap B = \emptyset$.

Informally, D-consistency ensures that any two quorums must intersect in a correct server. D-availability ensures the existence of a quorum containing only correct servers.

▶ **Example 1.** Consider a universe **V** with $3f + 1$ servers, a quorum system $\mathcal{Q}$ that contains every set with at least $2f + 1$ servers, and a fail-prone system $\mathcal{B}$ that contains every set with exactly $f$ servers. Then $(\mathcal{Q}, \mathcal{B})$ is a DQS. Indeed, D-consistency holds because any two quorums have at least $f + 1$ servers in common, at least one of which must be correct. D-availability holds because, for each $B \in \mathcal{B}$, the set $\mathbf{V} \setminus B$ has $2f + 1$ servers and is thus a quorum in $\mathcal{Q}$.

▶ **Example 2.** Consider the universe $\mathbf{V} = \{1, 2, 3, 4\}$, the quorum system

$$\mathcal{Q} = \{\{1, 2\}, \{1, 2, 3\}, \{1, 3, 4\}, \{1, 2, 3, 4\}\}$$

and the fail-prone system $\mathcal{B} = \{\{2\}, \{3, 4\}\}$. Then $(\mathcal{Q}, \mathcal{B})$ is a DQS. D-consistency holds because all quorums in $\mathcal{Q}$ intersect at 1, which does not belong to any element of $\mathcal{B}$. D-availability holds because both $\{1, 3, 4\}$ and $\{1, 2\}$ are quorums.

**Bracha broadcast over DQS.**   We next consider the abstraction of *reliable Byzantine broadcast* and show that a generalisation of Bracha broadcast [2] implements it when executed over an arbitrary dissemination quorum system. This protocol is the closest one from the world of classical BQSes that matches the broadcast protocol used in Stellar, and we relate the two in the following sections.

```
 1  process client(c ∈ C)
 2  │  broadcast(a)
 3  │  └  send BCAST(a) to every v ∈ V;

 4  process server(v ∈ V)
 5  │  echoed, ready, delivered ← ff ∈ Bool;
 6  │  when received BCAST(a) from c and echoed = ff
 7  │  │  echoed ← tt;
 8  │  └  send ECHO(a) to every v' ∈ V;
 9  │  when received ECHO(a) from every u ∈ U for some U ∈ Q and ready = ff
10  │  │  ready ← tt;
11  │  └  send READY(a) to every v' ∈ V;
12  │  when received READY(a) from every u ∈ B for some B ∈ 2^V \ {∅} such
    │     that ∀B' ∈ B. B ⊈ B' and ready = ff
13  │  │  ready ← tt;
14  │  └  send READY(a) to every v' ∈ V;
15  │  when received READY(a) from every u ∈ U for some U ∈ Q and
    │     delivered = ff
16  │  │  delivered ← tt;
17  └  └  deliver(a);
```

■ **Figure 1** Bracha broadcast over an arbitrary DQS $(Q, B)$.

The broadcast allows a distinguished client to broadcast a value $a$ to a group of servers, an event which we denote broadcast($a$). We denote the event of a server delivering a value $a$ as deliver($a$). Both the sender and some of the receivers can be Byzantine. *Reliable Byzantine broadcast* is defined by the following properties [3]:

- *Validity:* If the sender client is correct and broadcasts a value $a$, then every correct server eventually delivers $a$.
- *No duplication:* Every correct server delivers at most one value.
- *Integrity:* If some correct server delivers a value $a$ and the sender client is correct, then $a$ was previously broadcast by the sender.
- *Consistency:* If some correct server delivers a value $a$ and another correct server delivers a value $a'$, then $a = a'$.
- *Totality:* If a correct server delivers a value, then every correct server eventually delivers a value.

The Validity property ensures that the broadcast operates as expected when the sender is correct. When the sender is faulty, the above properties guarantee that servers may not deliver contradictory values, even if the sender sends different values to different servers. In this case they may also not deliver any value at all, but the liveness property of Totality ensures that all correct servers behave the same in this respect.

In Figure 1 we present an implementation of reliable Byzantine broadcast over a DQS $(Q, B)$. This generalises *Bracha's protocol* [2] for reliable Byzantine broadcast, which works over the cardinality-based DQS from Example 1. In order to broadcast a value $a$, the client sends a `BCAST(a)` message to every server (line 2). Servers process client messages in several phases, with progress denoted by several Boolean flags (line 5).

Since a faulty client may send contradictory messages to different servers, to ensure Consistency the servers first cross-check the client's proposals. A server that receives a message `BCAST(a)` for the first time sends an `ECHO(a)` message to all servers (including itself, for uniformity; line 6). When the server receives a quorum of `ECHO(a)` messages, it sends a `READY(a)` message to every server, signalling its willingness to deliver the value $a$ (line 9). Note that correct servers cannot send `READY` messages in this way with two different values: this would require two quorums of `ECHO` messages with different values, but by D-consistency these quorums would intersect in a correct server, which cannot send contradictory `ECHO` messages. When a server receives a quorum of `READY(a)` messages, it delivers $a$ (line 15).

The exchange of `READY` messages is necessary to ensure Totality: it guarantees that, if a correct server delivers a value, other correct servers have enough information to also deliver the value. This relies on an additional rule in line 12, allowing on a server to send a `READY(a)` message even if it previously echoed a different value: this is done if the server receives `READY(a)` from each member of a set $B$ that is not a subset of any element of $\mathcal{B}$. In this case at least one of the servers in $B$ must be correct, and hence, we can trust that the value it proposes has been cross-checked for consistency by a quorum.

▶ **Theorem 3.** *Let $(\mathcal{Q}, \mathcal{B})$ be a DQS. Then the protocol in Figure 1 satisfies the specification of reliable Byzantine where all faulty servers belong to some element of $\mathcal{B}$.*

The next example shows that removing the handler at line 12 of Figure 1 would undermine Totality.

▶ **Example 4.** Consider the DQS $(\mathcal{Q}, \mathcal{B})$ from Example 1, where quorums and fail-prone sets are defined by their cardinalities, and let $f = 1$ with a total number of 4 servers. As before, assume servers 1, 2 and 4 are correct, and 3 is faulty. Let the sender be faulty and send contradictory `BCAST(a)` and `BCAST(a')` messages to the sets $\{1, 2\}$ and $\{4\}$, respectively. Imagine faulty 3 sends `ECHO(a)` to 1 and 2 and never sends anything to 4. Then both 1 and 2 will send `READY(a)` to every server thanks to the conditions in line 9. If now 3 sends `READY(a)` to 1 and 2, then both will deliver value $a$ thanks to the conditions in line 15. However, correct 4 cannot send `READY(a)` and thus cannot deliver $a$, because it never receives `ECHO(a)` from a quorum. But correct 1 and 2 did deliver $a$. By enabling the handler at line 12, 4 would send `READY(a)` to every server after receiving `READY(a)` from 1 and 2, because any set of two servers is not contained in any of the elements of $\mathcal{B}$, which are singletons. After receiving `READY(a)` from 1, 2 and 4, server 4 will also deliver $a$ thanks to the conditions in line 15.

The next example demonstrates that delivering values right after receiving a quorum of `ECHO` messages would also undermine Totality, thus motivating the need for `READY` messages.

▶ **Example 5.** Consider the DQS $(\mathcal{Q}, \mathcal{B})$ and the set of faulty servers from Example 4 and assume that the sender is faulty and sends contradictory `BCAST(a)` and `BCAST(a')` messages to the sets $\{1\}$ and $\{2, 4\}$, respectively. Imagine 3 sends `ECHO(a')` to 2. Then 2 receives `ECHO(a')` from the quorum $\{2, 3, 4\}$. Imagine now that the faulty server 3 stops sending messages. If servers delivered values after receiving a quorum of `ECHO` messages, then correct server 2 would deliver $a'$, but correct servers 1 and 4 would never deliver any value.

## 3    Federated Byzantine Quorum Systems

**FBQS definition.** We now consider *federated Byzantine quorum systems (FBQS)*, which are used in the Stellar blockchain to construct quorums from the trust choices of individual participants [10]. An FBQS is a function $\mathcal{S} : \mathbf{V} \to 2^{2^{\mathbf{V}}} \setminus \{\emptyset\}$ that specifies the set of *quorum slices* for each server, ranged over by $q$. We require that a server belong to all of its own quorum slices: $\forall v \in \mathbf{V}. \forall q \in \mathcal{S}(v). v \in q$. Quorum slices reflect the trust choices of each server: a server can be convinced to accept the validity of a statement by any of its slices.

An FBQS specifies quorum slices for all servers. However, to participate in a protocol such as reliable broadcast or consensus, a server may not need to know the whole FBQS, which is infeasible in a system with open membership. Instead, it may be sufficient for the server to know only the quorum slices for the servers it interacts with, which it can receive directly from these servers. A complication is that faulty servers may lie about their quorum slices and, in particular, give contradictory information to their peers. In this section we consider an idealised setting (used in the Stellar whitepaper [10]) that assumes this does not happen: faulty servers do not equivocate about their quorum slices, even though they can otherwise behave arbitrarily. Hence, every server knows a part of the same FBQS $\mathcal{S}$. We lift this assumption in §5.

A non-empty set of servers $U \subseteq \mathbf{V}$ is a *quorum* in an FBQS $\mathcal{S}$ if $U$ contains a slice for each member: $\forall v \in U. \exists q \in \mathcal{S}(v). q \subseteq U$. Note that a server can check whether $U$ is a quorum based solely on the quorum slices of its participants. Hence, a server can discover quorums without knowing a complete FBQS by probing different servers until it finds a set forming a quorum. Alternatively, (signed) information about quorum slices of a given server can be transitively propagated via chains of trust dependencies, with each server receiving information from the servers in its slices. We say that an FBQS enjoys *quorum intersection* when any two of its quorums intersect. An FBQS $\mathcal{S}$ with quorum intersection induces a quorum system $\mathcal{Q}$ consisting of its quorums.

▶ **Example 6.** Consider an FBQS with $3f + 1$ servers where every server has a slice for each set of $2f + 1$ servers. This FBQS induces the quorum system $\mathcal{Q}$ from from Example 1: any set of $2f + 1$ or more servers is a quorum.

▶ **Example 7.** Consider the following FBQS, where each server has only one slice determined by outgoing arrows, except server 1, which has two slices respectively determined by the solid and the dashed arrows (we omit self-loops in the diagram to avoid clutter). This FBQS induces the quorum system $\mathcal{Q}$ from Example 2.

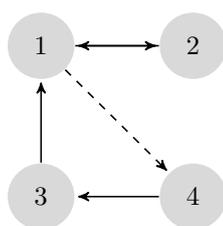

$$\begin{aligned}
\mathcal{S}(1) &= \{\{1,2\},\{1,4\}\} \\
\mathcal{S}(2) &= \{\{1,2\}\} \\
\mathcal{S}(3) &= \{\{1,3\}\} \\
\mathcal{S}(4) &= \{\{3,4\}\}
\end{aligned}$$

**Intact servers.** The correctness of protocols over FBQSes is stated using the following notion of *intact* servers [10]. Informally, the set of intact servers is the biggest one that excludes all faulty servers and still yields a quorum system that satisfies certain properties allowing correct protocol operation within this set. Formally, consider an FBQS $\mathcal{S}$ and a set of servers $I$. We define the *projection* $\mathcal{S}|_I$ of $\mathcal{S}$ to $I$ as the FBQS over universe $I$ defined as follows: $\mathcal{S}|_I(v) = \{q \cap I \mid q \in \mathcal{S}(v)\}$. Given a set of faulty servers $\mathbf{V}_{\text{bad}}$, the set of *intact* servers $\mathbf{V}_{\text{int}}$ is the biggest set such that

- all servers in $\mathbf{V}_{\text{int}}$ are correct: $\mathbf{V}_{\text{int}} \cap \mathbf{V}_{\text{bad}} = \emptyset$;
- if $\mathbf{V}_{\text{int}}$ is non-empty, then it is a quorum in $\mathcal{S}$;

= $\mathcal{S}|_{\mathbf{V}_{\mathrm{int}}}$ enjoys quorum intersection.

We call the servers in $\mathbf{V} \setminus \mathbf{V}_{\mathrm{int}}$ *befouled*. Informally, even though befouled servers may be correct, they "trust wrong guys" and thus protocols we consider may not guarantee all correctness properties for them.

▶ **Proposition 8.** *The set of intact servers in an FBQS with quorum intersection is well-defined.*

For the FBQS from Example 6, which corresponds to a cardinality-based DQS, the notions of correct and intact coincide. However, this is not the case for the FBQS $\mathcal{S}$ from Example 7. Namely, assume servers 1, 2 and 4 are correct and server 3 is faulty. The set $\{1, 2\}$ is a quorum in $\mathcal{S}$, and $\mathcal{S}|_{\{1,2\}}$ satisfies quorum intersection. Furthermore, adding the correct server 4 to the set $\{1, 2\}$ does not result in a quorum in $\mathcal{S}$. Hence, servers 1 and 2 are intact and server 4 is correct, but befouled. Intact servers satisfy the following useful property [10], which is similar to the D-consistency property of DQSes.

▶ **Proposition 9.** *Consider an FBQS $\mathcal{S}$ with quorum intersection. Fix a set of faulty servers and assume that there exists at least one intact server in $\mathcal{S}$. Then the intersection of every two quorums in $\mathcal{S}$ contains some intact server.*

**From federated to classical quorum systems.** We now show that every FBQS induces a corresponding DQS. An FBQS does not explicitly specify the possible failure scenarios, since in a setting with open membership one cannot easily predict them. Hence, to construct a fail-prone system for an FBQS, we consider all possible failures that do not bring the whole system down, i.e., leave at least some servers intact.

▶ **Theorem 10.** *Consider an FBQS $\mathcal{S}$ with quorum intersection. Let $\mathcal{Q}$ be the quorum system it induces and let $\mathcal{B}$ consist of the maximal sets $B$ such that the failure of $B$ leaves the set of intact servers in $\mathcal{S}$ non-empty. Then $(\mathcal{Q}, \mathcal{B})$ is a DQS.*

We say that FBQS $\mathcal{S}$ *induces* the DQS $(\mathcal{Q}, \mathcal{B})$ from the theorem. For example, the FBQS $\mathcal{S}$ from Example 6 induces the DQS from Example 1. Similarly, the FBQS $\mathcal{S}$ from Example 7 induces the DQS from Example 2. Indeed, the failure of server 2 leaves 1, 3 and 4 intact, and the failure of 3 and 4 leaves 1 and 2 intact. Extending either of these failure sets leaves no servers intact. The correspondence stated by Theorem 10 allows us to run any off-the-shelf protocol over a DQS on top of an FBQS with at least one intact server. In particular, the following corollary of Theorems 3 and 10 shows that this is the case for Bracha broadcast.

▶ **Corollary 11.** *Let $\mathcal{S}$ be an FBQS with quorum intersection and $(\mathcal{Q}, \mathcal{B})$ be the DQS it induces. Then the protocol in Figure 1 satisfies the specification of reliable Byzantine broadcast in executions where at least one server in $\mathcal{S}$ is intact.*

Note that, unlike in Theorem 3, where the assumption on the allowed failure scenarios was stated using a given $\mathcal{B}$, here this is done by requiring the existence of at least one intact server. For example, consider the FBQS from Example 6. The existence of an intact server in this FBQS is equivalent to requiring that no more than $f$ servers fail, which corresponds to the standard requirement for the cardinality-based DQS (Example 1).

## 4    Stellar Broadcast and its Correctness

**Stellar broadcast.** Despite Corollary 11, using Bracha broadcast in a federated setting is problematic. While servers can discover quorums in $\mathcal{Q}$ without knowing the whole FBQS,

Bracha broadcast also requires participants to know the fail-prone system $\mathcal{B}$ (line 12). Computing the $\mathcal{B}$ induced by $\mathcal{S}$ requires knowing the quorum slices of all participants, which is infeasible in a system with open membership. Stellar solves this problem by replacing the condition in line 12 by one that a server can check locally. We now present and prove correct the corresponding broadcast protocol, which is a reformulation of Stellar's *federated voting* protocol; we call the resulting protocol the *Stellar broadcast*.

Given an FBQS $\mathcal{S}$ and a server $v$, a set of servers $B$ is called $v$-*blocking* if $B$ overlaps each quorum slice in $\mathcal{S}(v)$. The Stellar broadcast is obtained from Bracha broadcast in Figure 1 by making the following changes to three lines:

> 9: **when received** ECHO($a$) **from each** $u \in U$ **for some** $U \in \mathcal{Q}$ ***such that*** $v \in U$
>    **and** ready $= f\!f$;
> 12: **when received** READY($a$) **from every** $u \in B$ ***for some*** $v$-***blocking*** $B \in 2^{\mathbf{V}} \setminus \{\emptyset\}$
>    **and** ready $= f\!f$;
> 15: **when received** READY($a$) **from every** $u \in U$ **for some** $U \in \mathcal{Q}$
>    ***such that*** $v \in U$ **and** delivered $= f\!f$,

where $v$ is the server executing the code. The new checks at lines 9 and 15 require a server to only accept information from quorums it belongs to. The new check at line 12 depends only on the slices of $v$ and can thus be performed locally. This allows $v$ to operate even when knowing only the part of the FBQS corresponding to the servers it directly interacts with. The role of the new check in line 12 resembles that of the old check in Figure 1. The old check guaranteed that the server $v$ has received at least one READY($a$) message from a *correct* server. Due to the following proposition, the new check guarantees that, *if $v$ is intact*, then it has received at least one READY($a$) message from an *intact* server.

▶ **Proposition 12.** *Assume an FBQS $\mathcal{S}$ with quorum intersection. For a given set of faulty servers, the set $\mathbf{V}_{\text{bef}}$ of befouled servers is not $v$-blocking for any intact $v$.*

**Proof.** Consider an intact server $v$. The set of all intact servers $\mathbf{V}_{\text{int}} = \mathbf{V} \setminus \mathbf{V}_{\text{bef}}$ is a quorum in $\mathcal{S}$. Then there exists a slice $q \in \mathcal{S}(v)$ such that $q \subseteq \mathbf{V}_{\text{int}}$. Hence, $\mathbf{V}_{\text{bef}}$ cannot overlap all of $v$'s slices, as required. ◀

As it happens, the Stellar broadcast does not implement reliable Byzantine broadcast, but only its weaker version, which we call *weakly reliable Byzantine broadcast*. This weakens the properties of Validity and Totality (§2) by guaranteeing them only to intact servers:

- *Validity for intact servers:* If the sender client is correct and broadcasts a value $a$, then every intact server eventually delivers $a$.
- *Totality for intact servers:* If a correct server delivers a value, then every intact server eventually delivers a value.

As we noted above, for the cardinality-based quorum system from Example 1 the notions of intact and correct coincide. Hence, for this quorum system the weak and the classical broadcast abstractions coincide as well.

▶ **Theorem 13.** *Let $\mathcal{S}$ be an FBQS with quorum intersection. The protocol in Figure 1 over $\mathcal{S}$ with the changed lines 9, 12 and 15 satisfies the specification of weakly reliable Byzantine broadcast in executions where at least one server in $\mathcal{S}$ is intact.*

As the next example illustrates, the Stellar broadcast does not satisfy the Totality property of reliable Byzantine broadcast.

▶ **Example 14.** Consider the FBQS $\mathcal{S}$ from Example 7. Assume the sender is faulty and it sends contradictory BCAST($a$) and BCAST($a'$) messages to the sets $\{1, 2\}$ and $\{4\}$ respectively. Servers 1 and 2 will receive ECHO($a$) from the quorum $\{1, 2\}$, and by the conditions in the changed lines 9 and 15, they will then send READY($a$) to every server and deliver value $a$. However, 4 will not be able to deliver $a$, because it will receive READY($a$) from the set $\{1, 2\}$, which is a quorum to which 4 does not belong.

The condition in the Stellar broadcast that a server only accepts information from quorums it belongs to (lines 9 and 15) is not strictly required to allow a server to operate without the knowledge of the whole system: a server can ensure that a set of servers is a quorum based on the quorum slices of its participants, regardless of whether the server is one of them. If we drop the above condition, then the protocol implements the usual reliable Byzantine broadcast.

▶ **Theorem 15.** *Let $\mathcal{S}$ be an FBQS with quorum intersection. The protocol in Figure 1 over $\mathcal{S}$ with the changed line 12 satisfies the specification of reliable Byzantine broadcast in executions where at least one server in $\mathcal{S}$ is intact.*

The above theorems yield several novel insights with respect to the existing results about Stellar. First, they show that some of the the ad hoc protocols proposed for Stellar actually implement a variant of a well-known broadcast abstraction. Second, the theorems strengthen the existing correctness theorems for Stellar's federated voting [10]. Namely, we guarantee Consistency for all correct servers, whereas the Consistency property proved for federated voting [10, Theorem 9] applies only to a subset of correct servers $I$ such that $\mathcal{S}|_I$ enjoys quorum intersection. The difference is significant in practice, since a server cannot easily find out if it belongs to such a subset. The Totality property proved for federated voting [10, Theorem 11] requires an intact server to deliver a value, whereas we show that a correct server delivering a value is enough for all intact servers to also deliver the value. Our theorems also show that, assuming the existence of at least one intact server, the guarantees provided by the protocol can be strengthened if a server accepts information from quorums it does not participate in. In §5 we strengthen our correctness statements even further by considering the case when faulty servers may equivocate about their quorum slices.

**Proving correctness.** We give the key steps in proving Theorem 13, which we revisit in §5 to show that the protocol also works when faulty servers equivocate about their quorum slices. The following lemma is used to establish Consistency.

▶ **Lemma 16.** *Assume an FBQS $\mathcal{S}$ with quorum intersection. If intact servers in an execution of Stellar broadcast over $\mathcal{S}$ send messages* READY($a$) *and* READY($a'$)*, then $a = a'$.*

*Proof sketch.* To prove this lemma, using Proposition 12 we establish that, if intact servers send messages READY($a$) and READY($a'$), then some (possibly different) intact servers receive quorums of messages ECHO($a$) and ECHO($a'$), respectively. By Proposition 9 these quorums have to intersect in an intact server, which is also correct and thus cannot send contradictory ECHO messages.                                                                                           ◀

*Proof sketch for Consistency.* Assume that correct servers deliver values $a$ and $a'$. Then by the condition in the changed line 15, the corresponding servers received a quorum of READY($a$) and READY($a'$) messages, respectively. Since at least one server in $\mathcal{S}$ is intact, the set of intact servers forms a quorum. Then, since $\mathcal{S}$ has quorum intersection, each of the quorums that sent the READY($a$) and READY($a'$) messages must contain at least one intact server. But then by Lemma 16 we have $a = a'$, as required.                                                                                           ◀

The next lemma is used to prove Totality.

▶ **Lemma 17.** *Let $\mathcal{S}$ be an FBQS with quorum intersection. Consider an execution of the Stellar broadcast with the set of intact servers in $\mathcal{S}$ being $\mathbf{V}_{\text{int}} \neq \emptyset$. Assume that $\mathbf{V}_{\text{int}} = V^+ \uplus V^-$ and for some quorum $U$ we have $U \cap \mathbf{V}_{\text{int}} \subseteq V^+$. Then either $V^- = \emptyset$ or there exists some server $v \in V^-$ such that $V^+$ is v-blocking.*

*Proof sketch for Totality.* Delivering a value $a$ at an intact server requires this server to receive $\texttt{READY}(a)$ messages from a quorum $U$. Since the set of intact servers forms a quorum and any two quorums intersect, the quorum $U$ must contain some intact servers. Using the above lemma, we can then show that, as the $\texttt{READY}(a)$ messages from these intact servers propagate through the system, more and more intact servers will execute the handler at line 12 and send $\texttt{READY}(a)$ messages themselves. When applying the lemma, the set $V^+$ is the set of intact servers that have sent $\texttt{READY}(a)$ and $V^-$ the set of intact servers that have not done so yet. The lemma ensures that in the end all intact servers will send $\texttt{READY}(a)$ and will be able to deliver $a$.                                                                                ◀

**Observational equivalence.** We now establish a tighter correspondence between the Stellar broadcast over an FBQS and Bracha broadcast over the corresponding BQS: we show that Bracha broadcast is observationally equivalent to the version of the Stellar broadcast with the changed line 12 only. Informally, this means that any externally observable behaviour of one of the protocols can also be produced by the other. Note that this relationship does not simply follow from the fact that the two broadcasts implement the same specification, because the specification is non-deterministic: when the sender is faulty and sends contradictory values, it allows receivers to pick one value to deliver, or not to deliver any value at all. The equivalence thus shows that, whenever servers in one of the protocols decide to deliver a value, there is an execution of the other protocol (with some behaviour of faulty servers) that also delivers the same value.

We formulate the equivalence using the following notion. A *history* $H$ is a set of events in an execution of a broadcast protocol that includes the reception of the first $\texttt{BCAST}$ message received by each server and the triggering of the deliver primitive by each server. A history captures the initial communication between the clients and the system and the outcome of the protocol's execution that is observable by applications running at the servers.

▶ **Theorem 18.** *Let $\mathcal{S}$ be an FBQS with quorum intersection. Fix the sets of faulty and correct servers such that at least one intact server exists in $\mathcal{S}$, and let $(\mathcal{Q}, \mathcal{B})$ be the DQS that $\mathcal{S}$ induces.*

  **(i)** *For every execution of Bracha broadcast over $(\mathcal{Q}, \mathcal{B})$ with a history $H$, there exists an execution of the Stellar broadcast over $\mathcal{S}$ with the changed line 12 that also entails $H$.*
  **(ii)** *For every execution of the Stellar broadcast over $\mathcal{S}$ with the changed line 12 that has a history $H$, there exists an execution of Bracha broadcast over $(\mathcal{Q}, \mathcal{B})$ with a history $H$.*

## 5 Lying about Trust Choices

So far we have assumed that faulty servers do not equivocate about their quorum slices, so that all servers share the same FBQS $\mathcal{S}$. This is unrealistic, since in practice servers find out about quorum slices of their peers from the peers themselves, and faulty peers may lie. This may lead different servers to have inconsistent information about the quorum slices of a given faulty server. To capture this, we generalise the notion of an FBQS to allow different servers to have different views on quorum slices. Let us fix the sets of correct and faulty servers, $\mathbf{V}_{\mathrm{ok}}$ and $\mathbf{V}_{\mathrm{bad}}$. We say that an indexed family $\{\mathcal{S}_v\}_{v \in \mathbf{V}_{\mathrm{ok}}}$ is a *subjective FBQS* if each $\mathcal{S}_v$ is an FBQS and the different FBQSes agree on the quorum slices of correct servers: $\forall v_1, v_2, v \in \mathbf{V}_{\mathrm{ok}}. \mathcal{S}_{v_1}(v) = \mathcal{S}_{v_2}(v)$. A subjective FBQS reflects each correct server's view on the choices of trust of other servers. We thus sometimes call $\mathcal{S}_v$ the *view* of the server $v$. We say that $q$ is $u$'s slice *known by* $v$ if $q \in \mathcal{S}_v(u)$. Note that a subjective FBQS does not specify the views of faulty servers, as they are immaterial.

The fact that different servers may have inconsistent information about quorum slices means that the same holds for quorums. Given a subjective FBQS $\{\mathcal{S}_v\}_{v \in \mathbf{V}_{\mathrm{ok}}}$, we say that $U$ is a *quorum known by* $v$ when $U$ is a quorum in FBQS $\mathcal{S}_v$. As before, a server can discover (subjective) quorums by asking its peers for their slices until it finds a set forming a quorum. A subjective FBQS $\{\mathcal{S}_v\}_{v \in \mathbf{V}_{\mathrm{ok}}}$ satisfies *quorum intersection* if $\mathcal{S}_v$ satisfies quorum intersection for every $v$ that is correct. A subjective FBQS $\{\mathcal{S}_v\}_{v \in \mathbf{V}_{\mathrm{ok}}}$ with quorum intersection induces a *subjective quorum system* $\{\mathcal{Q}_v\}_{v \in \mathbf{V}_{\mathrm{ok}}}$, consisting of the quorums known by each correct server.

▶ **Example 19.** Let $\mathbf{V} = \{1, 2, 3, 4\}$ and assume that 3 is faulty and the rest are correct. Consider a subjective FBQS defined by the following diagrams:

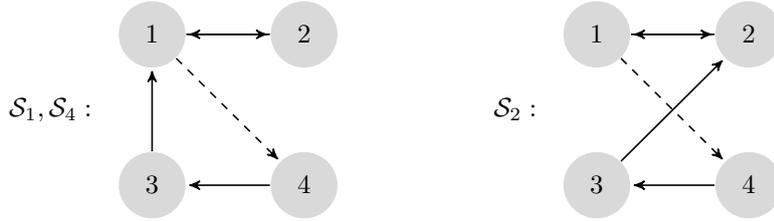

Each server's slices are determined by the outgoing arrows (we omit self-loops). In the case of server 1 we have the two slices determined respectively by the solid and the dashed arrows. Hence, we have

$$\mathcal{S}_1(1) = \mathcal{S}_2(1) = \mathcal{S}_4(1) = \{\{1, 2\}, \{1, 4\}\}; \qquad \mathcal{S}_1(2) = \mathcal{S}_2(2) = \mathcal{S}_4(2) = \{\{1, 2\}\};$$
$$\mathcal{S}_1(4) = \mathcal{S}_2(4) = \mathcal{S}_4(4) = \{\{3, 4\}\}.$$

The faulty server 3 communicates the slices $\mathcal{S}_1(3) = \mathcal{S}_4(3) = \{\{1, 3\}\}$ to servers 1 and 4, and the slices $\mathcal{S}_2(3) = \{\{2, 3\}\}$ to server 2. The above FBQS induces a subjective quorum system with

$$\mathcal{Q}_1 = \mathcal{Q}_4 = \{\{1, 2\}, \{1, 2, 3\}, \{1, 3, 4\}, \{1, 2, 3, 4\}\}; \quad \mathcal{Q}_2 = \{\{1, 2\}, \{1, 2, 3\}, \{1, 2, 3, 4\}\}.$$

Thus, due to server 3 equivocating, servers 1 and 4 consider $\{1, 3, 4\}$ a quorum, but server 2 does not.

We next adapt the notion of intact servers to subjective FBQSes by requiring the conditions satisfied by intact servers to hold in any subjective view. Given a subjective FBQS $\{\mathcal{S}_v\}_{v \in \mathbf{V}_{\mathrm{ok}}}$, the set of *intact servers* $\mathbf{V}_{\mathrm{int}}$ is the biggest set such that

- all servers in $\mathbf{V}_{\mathrm{int}}$ are correct: $\mathbf{V}_{\mathrm{int}} \cap \mathbf{V}_{\mathrm{bad}} = \emptyset$;
- if $\mathbf{V}_{\mathrm{int}}$ is non-empty, then it is a quorum in $\mathcal{S}_v$ for each $v \in \mathbf{V}_{\mathrm{ok}}$;
- for each $v \in \mathbf{V}_{\mathrm{ok}}$, $\mathcal{S}_v|_{\mathbf{V}_{\mathrm{int}}}$ enjoys quorum intersection.

Note that the definition of intact servers depends only on the slices of correct servers in $\{\mathcal{S}_v\}_{v \in \mathbf{V}_{\mathrm{ok}}}$, which agree across the views $\mathcal{S}_v$ of different servers $v$. Hence, the notions of intact computed from the subjective views of every server coincide with the above notion.

▶ **Proposition 20.** *The set of intact servers in a subjective FBQS $\{\mathcal{S}_v\}_{v \in \mathbf{V}_{ok}}$ with quorum intersection is well-defined and is equal to the set of intact servers in any of $\mathcal{S}_v$.*

For instance, the set of intact servers in any of the subjective views in Example 19 is $\{1, 2\}$. As before, servers that are not intact are called *befouled*.

Using the fact that the views of different servers agree on the quorum slices of correct servers, we can prove the following proposition, which ensures that the intersection of any two quorums, even when coming from the views of different servers, has to contain an intact node. This proposition generalises Proposition 9 and is key for proving the correctness of the Stellar broadcast in the subjective setting.

▶ **Proposition 21.** *Let $\{\mathcal{S}_v\}_{v \in \mathbf{V}_{ok}}$ be a subjective FBQS with quorum intersection and at least one intact server, and let $\{\mathcal{Q}_v\}_{v \in \mathbf{V}_{ok}}$ be its induced subjective quorum system. Then the intersection of every two quorums in $\bigcup_{v \in \mathbf{V}_{ok}} \mathcal{Q}_v$ contains some intact server.*

For example, any two quorums in the quorum system from Example 19 intersect at 1, which is intact.

We can run the Stellar broadcast (with the changes to lines 9, 12 and 15; §4) over a subjective FBQS $\{\mathcal{S}_v\}_{v \in \mathbf{V}_{ok}}$ by letting each correct server $v$ act according to its FBQS $\mathcal{S}_v$, which it constructs based on the information provided by its peers. In particular, a set $B \subseteq \mathbf{V}$ is now considered $v$-blocking when it overlaps every slice in $\mathcal{S}_v(v)$: $\forall q \in \mathcal{S}_v(v). q \cap B \neq \emptyset$. In this setting, the Stellar broadcast still implements weakly reliable Byzantine broadcast, defined as before using the above notion of intact.

▶ **Theorem 22.** *Given sets of correct and faulty servers, $\mathbf{V}_{ok}$ and $\mathbf{V}_{bad}$, let $\{\mathcal{S}_v\}_{v \in \mathbf{V}_{ok}}$ be a subjective FBQS with quorum intersection and at least one intact server. Then the Stellar broadcast over this system implements weakly reliable Byzantine broadcast.*

We can also establish an analogue of Theorem 15, showing that, when servers accept quorums they do not belong to, the broadcast guarantees are strengthened.

The above theorem is our key result: it shows that the Stellar broadcast is correct in the setting of its intended deployment, where servers may have inconsistent information about the trust choices of their peers. This setting has not been considered by the existing correctness theorems for Stellar [10]. Note that, like in the idealised setting we considered earlier, a server in the Stellar broadcast is able to operate when knowing only the quorum slices of the servers it directly interacts with (even though those reported by faulty servers may be bogus). Theorem 22 holds because the key lemmas used to prove the correctness of the Stellar broadcast in the idealised setting can be adjusted to the more general setting we consider here. The following lemma generalises Lemmas 16 and is used to prove Consistency.

▶ **Lemma 23.** *Assume a subjective FBQS $\{\mathcal{S}_v\}_{v \in \mathbf{V}_{ok}}$ with quorum intersection and consider an execution of Stellar broadcast over $\{\mathcal{S}_v\}_{v \in \mathbf{V}_{ok}}$. If intact servers send messages `READY`$(a)$ and `READY`$(a')$, then $a = a'$.*

We prove the above lemma similarly to Lemma 16, but using Proposition 21 instead of Proposition 9. We prove Totality similarly to the idealised case using the following lemma, generalising Lemma 17.

▶ **Lemma 24.** *Let $\{\mathcal{S}_v\}_{v \in \mathbf{V}_{ok}}$ be a subjective FBQS with quorum intersection. Consider an execution of the Stellar broadcast with the set of intact servers in $\{\mathcal{S}_v\}_{v \in \mathbf{V}_{ok}}$ being $\mathbf{V}_{int} \neq \emptyset$. Assume that $\mathbf{V}_{int} = V^+ \uplus V^-$ and for some quorum $U$ known by a server $v$ we have $U \cap \mathbf{V}_{int} \subseteq V^+$. Then either $V^- = \emptyset$ or there exists some server $v' \in V^-$ such that $V^+$ is $v'$-blocking.*

## 6    Subjective Dissemination Quorum Systems

We have shown that a federated system where servers may lie about their trust choices generates a system in which every server has its own subjective notion of a quorum. We now investigate such subjective quorum structures independently from the way they are generated, which may be useful in the future for dealing with approaches to decentralising trust different from Stellar. To this end, we introduce the notion of a *subjective Byzantine quorum system*, where different servers may have different Byzantine quorum systems.

Let us fix the sets of correct and faulty servers, $\mathbf{V}_{\text{ok}}$ and $\mathbf{V}_{\text{bad}}$. A *subjective Byzantine quorum system* is a pair of two indexed families—a subjective quorum system $\{\mathcal{Q}_v\}_{v \in \mathbf{V}_{\text{ok}}}$ and a *subjective fail-prone system* $\{\mathcal{B}_v\}_{v \in \mathbf{V}_{\text{ok}}}$, where each $\mathcal{B}_v$ is a fail-prone system. A subjective BQS $(\{\mathcal{Q}_v\}_{v \in \mathbf{V}_{\text{ok}}}, \{\mathcal{B}_v\}_{v \in \mathbf{V}_{\text{ok}}})$ is a *subjective DQS* when the following properties hold:

- *SD-safety:* $\forall v \in \mathbf{V}_{\text{ok}}. \exists B \in \mathcal{B}_v. \mathbf{V}_{\text{bad}} \subseteq B$;
- *SD-consistency:*
    $\forall v \in \mathbf{V}_{\text{ok}}. \forall U_1 \in \mathcal{Q}_v. \forall U_2 \in \bigcup_{v' \in \mathbf{V}_{\text{ok}}} \mathcal{Q}_{v'}. \forall B \in \mathcal{B}_v. \mathbf{V}_{\text{bad}} \subseteq B \implies U_1 \cap U_2 \nsubseteq B$; and
- *SD-availability:* each pair $(\mathcal{Q}_v, \mathcal{B}_v)$ with $v \in \mathbf{V}_{\text{ok}}$ satisfies D-availability.

SD-safety requires that the fail-prone system of each server contains some element that includes all faulty servers. Such a condition was not part of the previous DQS definition, but we required it when stating protocol correctness in a particular execution where the set of faulty servers was fixed (Theorem 3). Since a subjective DQS is already defined in the context of a fixed set of faulty servers, we add this condition to its definition. SD-consistency requires that the intersection of a quorum known by a server $v$ with a quorum known by any other server cannot be completely faulty according to $v$'s view. To formulate this we consider only sets $B \in \mathcal{B}_v$ that contain all faulty servers; at least one such set is guaranteed to exist by SD-safety. This requirement ensures that we can generate a subjective DQS from a subjective FBQS from a view of each correct server as per the mapping in Theorem 10.

▶ **Theorem 25.** *Given sets of correct and faulty servers, $\mathbf{V}_{\text{ok}}$ and $\mathbf{V}_{\text{bad}}$, let $\{\mathcal{S}_v\}_{v \in \mathbf{V}_{\text{ok}}}$ be a subjective FBQS with quorum intersection and with at least one intact server, and let $\{\mathcal{Q}_v\}_{v \in \mathbf{V}_{\text{ok}}}$ be the subjective quorum system it induces. For each $v \in \mathbf{V}_{\text{ok}}$ let $\mathcal{B}_v$ consist of the maximal sets $B$ such that the failure of $B$ leaves the set of intact servers in $\mathcal{Q}_v$ non-empty. Then $(\{\mathcal{Q}_v\}_{v \in \mathbf{V}_{\text{ok}}}, \{\mathcal{B}_v\}_{v \in \mathbf{V}_{\text{ok}}})$ is a subjective DQS.*

The properties of subjective DQSes are strong enough to execute Bracha broadcast assuming each correct server acts according to its view in an arbitrary subjective DQS.

▶ **Theorem 26.** *Given sets of correct and faulty servers, $\mathbf{V}_{\text{ok}}$ and $\mathbf{V}_{\text{bad}}$, let $(\{\mathcal{Q}_v\}_{v \in \mathbf{V}_{\text{ok}}}, \{\mathcal{B}_v\}_{v \in \mathbf{V}_{\text{ok}}})$ be a subjective DQS. Then Bracha broadcast over this system implements reliable Byzantine broadcast.*

In the special case when all views of correct servers are the same, a subjective DQS degenerates into a variant of a usual DQS $(\mathcal{Q}, \mathcal{B})$ that satisfies a weaker version of D-consistency that considers only sets $B \in \mathcal{B}$ that include all faulty servers:

$$\forall U_1, U_2 \in \mathcal{Q}. \forall B \in \mathcal{B}. \mathbf{V}_{\text{bad}} \subseteq B \implies U_1 \cap U_2 \nsubseteq B.$$

Theorem 26 implies that Bracha broadcast is also correct when executed over this DQS variant, which weakens the classical conditions required for correctness. The following stronger version of SD-consistency does specialise to D-consistency of DQSes:

$$\forall v \in \mathbf{V}_{\text{ok}}. \forall U_1 \in \mathcal{Q}_v. \forall U_2 \in \bigcup_{v' \in \mathbf{V}_{\text{ok}}} \mathcal{Q}_{v'}. \forall B \in \mathcal{B}_v. U_1 \cap U_2 \nsubseteq B.$$

Obviously, Bracha broadcast can also be correctly executed over a subjective DQS satisfying this property. Finally, Theorems 25 and 26 imply that Bracha broadcast can be executed over a subjective DQS induced by a subjective FBQS.

▶ **Corollary 27.** *Given sets of correct and faulty servers, $\mathbf{V}_{\mathrm{ok}}$ and $\mathbf{V}_{\mathrm{bad}}$, let $\{\mathcal{S}_v\}_{v \in \mathbf{V}_{\mathrm{ok}}}$ be a subjective FBQS with quorum intersection where at least one server is intact. Let $(\{\mathcal{Q}_v\}_{v \in \mathbf{V}_{\mathrm{ok}}}, \{\mathcal{B}_v\}_{v \in \mathbf{V}_{\mathrm{ok}}})$ be the subjective DQS that this FBQS induces. Then Bracha broadcast over this subjective DQS implements reliable Byzantine broadcast.*

Note that constructing the fail-prone system $\mathcal{B}_v$ required to run Bracha broadcast at server $v$ requires the server to have a complete (but possibly imperfect) information about the quorum slices of all servers in the system. Hence, the advantage of the Stellar broadcast over Bracha broadcast is preserved in the subjective setting.

## 7    Related Work

Byzantine quorum systems were proposed by Malkhi and Reiter [9], who also demonstrated how they can be used to implement read-write registers. The generalisation of Bracha broadcast to dissemination quorum systems in Figure 1 follows the techniques used in [9]. Researchers have also considered quorum systems that provide stronger guarantees than dissemination ones and studied which quorum systems yield the best performance characteristics [13]. An interesting avenue for future work is to investigate whether other types of quorum systems can be useful in the federated setting.

The Stellar broadcast as presented in this paper is based on the federated voting protocol, which is crucial to the Stellar consensus. In the Stellar whitepaper [10] this protocol is formulated as a form of binary consensus without termination guarantees. The whitepaper does not establish its relationship to reliable Byzantine broadcast or existing protocols. It does contain some correctness statements for the federated voting protocol; in particular, our proof of Totality (Lemma 17) follows the ideas in the whitepaper. However, the existing proofs only consider the idealised setting where Byzantine servers do not equivocate about their trust choices, which does not reflect the intended deployment of Stellar. They furthermore establish weaker safety and liveness properties than us.

## 8    Conclusions and Future Work

In this paper we have rigorously studied the basic concepts underlying the Stellar blockchain. In particular, we have established a formal correspondence between federated and classical Byzantine quorum systems, and between the federated voting protocol of Stellar and Bracha protocol for reliable Byzantine broadcast. Our results explicate the main benefit of designing a broadcast protocol tailored for the federated setting: allowing servers to operate based on incomplete information. By formalising the relationship between Stellar and Bracha broadcasts, we were additionally able to show that a variant of the Stellar broadcast closer to the original Bracha broadcast still allows servers to operate based on incomplete information, yet provides stronger guarantees.

We have also faithfully captured a realistic setting for deploying Stellar where Byzantine nodes may equivocate about their trust choices. We have shown that this setting motivates a generalisation of the classical Byzantine quorum systems that allows nodes to have inconsistent information about quorums. This may be useful in the future for dealing with alternative approaches to decentralising trust.

Our results demonstrate that the purpose of FBQSes is not limited to solving consensus, for which they were originally proposed. In particular, we believe that apart from broadcast abstractions, FBQSes can also be used to implement read-write registers on the lines of existing constructions [1, 9]. Even though in this paper we did not handle the whole of Stellar consensus protocol, in the future our results should also enable a rigorous analysis of this protocol and similar ones, such as Ripple [12]. Finally, we hope that the connections we have established between classical broadcast abstractions and the novel Stellar protocols will enable transferring ideas from the existing optimised protocols for BFT consensus [6, 8] into the federated setting.

**Acknowledgements.**    We thank Gregory Chockler for discussions about Byzantine quorum systems, and Ilya Sergey and Maria Schett for discussions about Stellar. This work was supported by an ERC Starting Grant RACCOON.

## A    Proofs for §2

▶ **Lemma 28.** *Consider a DQS* $(\mathcal{Q}, \mathcal{B})$ *and assume that some* $B \in \mathcal{B}$ *contains all the faulty servers. Then, every quorum in* $\mathcal{Q}$ *contains some correct server.*

**Proof.** By D-availability there is a quorum $U$ in $\mathcal{Q}$ that contains only correct servers. Consider an arbitrary quorum $U'$ in $\mathcal{Q}$. Since $\mathcal{Q}$ is a quorum system, the quorums $U$ and $U'$ intersect. Then, since $U$ contains only correct servers, $U'$ contains some correct server.    ◀

▶ **Lemma 29.** *Assume a DQS* $(\mathcal{Q}, \mathcal{B})$ *and consider an execution of the protocol in Figure 1 such that some* $B \in \mathcal{B}$ *contains all the faulty servers. The first correct server that sends a* READY($a$) *message first needs to receive* ECHO($a$) *messages from every member of a quorum.*

**Proof.** The READY($a$) message may be sent at lines 11 or 14. The case of line 14 is impossible: since some element of $\mathcal{B}$ contains all the faulty servers, a correct server cannot be the first to send a READY($a$) message at this line. Hence, READY($a$) is sent at line 11. Then the required follows from the condition in line 9.    ◀

▶ **Lemma 30.** *Assume a DQS* $(\mathcal{Q}, \mathcal{B})$ *and consider an execution of the protocol in Figure 1 such that some* $B \in \mathcal{B}$ *contains all the faulty servers. If a correct server delivers a value* $a$, *then some correct server received* ECHO($a$) *messages from every member of a quorum.*

**Proof.** Assume that some correct server delivers value $a$. Then this server must have received a quorum of READY($a$) messages. By Lemma 28, at least one correct server must have sent a READY($a$) message. By Lemma 29, the first server to do so received a quorum of ECHO($a$) messages.    ◀

▶ **Lemma 31.** *Assume a DQS* $(\mathcal{Q}, \mathcal{B})$ *and consider an execution of the protocol in Figure 1 such that some* $B \in \mathcal{B}$ *contains all the faulty servers. If correct servers send messages* READY($a$) *and* READY($a'$), *then* $a = a'$.

**Proof.** Assume that correct servers send messages READY($a$) and READY($a'$). Consider the first correct servers that send the respective messages, $v$ and $v'$. By Lemma 29, $v$ has received ECHO($a$) messages from a quorum $U$, and $v'$ has received ECHO($a'$) messages from a quorum $U'$. By D-consistency, $U \cap U'$ contains some correct server, so that this server has sent ECHO($a$) and ECHO($a'$). But due to the use of the guard variable echoed in lines 5 and 7, a server only echoes one value, and thus it cannot echo different values. Hence, $a = a'$, as required.    ◀

▶ **Lemma 32.** *Consider a DQS* $(\mathcal{Q}, \mathcal{B})$ *and assume that some element* $B \in \mathcal{B}$ *contains every faulty server. Let* $U$ *be a quorum in* $\mathcal{Q}$ *and let* $U^+$ *be the set of correct servers in* $U$. *Then, for every* $B \in \mathcal{B}$ *we have* $U^+ \nsubseteq B$.

**Proof.** Assume towards a contradiction that $U^+ \subseteq B$ for some $B \in \mathcal{B}$. By D-availability, $\mathcal{Q}$ contains a quorum $U' \subseteq \mathbf{V} \backslash U^+$. But then the $U \cap U'$ contains no correct server, contradicting D-consistency.    ◀

**Proof of Theorem 3.** *No duplication:* Straightforward by the use of the guard variable delivered in line 15.

*Integrity:* Assume that some correct server delivers value $a$. By Lemma 30, some correct server received a quorum of ECHO($a$) messages. Then by Lemma 28, at least one correct server sent ECHO($a$). Since the sender is correct, this is only possible if the sender broadcast $a$.

*Consistency:* Assume that some correct server delivers value $a$, and some correct server delivers value $a'$. Then by the condition in line 15, the servers received a quorum of `READY`($a$), respectively, `READY`($a'$) messages. By Lemma 28, either of these quorums contains some correct server. But then by Lemma 31 we have $a = a'$, as required.

*Validity:* Assume the sender is correct and broadcasts $a$. Then every correct server will eventually send `ECHO`($a$). By D-availability, there exists a quorum $U$ consisting only of correct servers, so that all the quorum members send `ECHO`($a$). Hence, every member of $U$ will eventually send `READY` messages due to the condition in line 9. By Lemma 31 these messages have to carry the value $a$. Then by the condition in line 15, every correct server will eventually deliver a value. Due to the Consistency property, this value has to be $a$.

*Totality:* Assume some correct server delivers value $a$. Then by the condition in line 15 the server has received `READY`($a$) messages from every member in a quorum $U$. Consider the set $U^+$ that contains the correct servers in $U$. By Lemma 28, $U^+ \neq \emptyset$. Every server in $U^+$ sends `READY`($a$) to every server. By Lemma 32, for every $B' \in \mathcal{B}$ we have $U^+ \not\subseteq B'$. Therefore, every correct server that has not sent `READY` messages already will eventually do so due to the condition in line 12. Due to Lemma 31, all these messages must carry the value $a$. Thus, every correct server will eventually send `READY`($a$). Since by D-availability there exists a quorum consisting only of correct servers, all correct servers will eventually deliver a value due to the condition in line 15.   ◄

## B   Proofs for §3

Let $\mathcal{S}$ be an FBQS and assume a set of faulty servers $\mathbf{V}_{\text{bad}}$. Let $I$ be a set of servers. We say that $\mathcal{S}$ enjoys

- *$I$-correctness:* iff all servers in $I$ are correct—*i.e.*, $I \cap \mathbf{V}_{\text{bad}} = \emptyset$;
- *$I$-availability:* iff either $I = \emptyset$ or $I$ is a quorum in $\mathcal{S}$; and
- *$I$-intersection:* iff $\mathcal{S}|_I$ enjoys quorum intersection.

According to the definition of intact from §3, the set of intact servers $\mathbf{V}_{\text{int}}$ is the biggest set $I$ such that $\mathcal{S}$ enjoys the three properties above.

▶ **Lemma 33.** *Let $U$ be a quorum in an FBQS $\mathcal{S}$, let $I$ be a set of servers, and let $U' = U \cap I$. If $U' \neq \emptyset$ then $U'$ is a quorum in $\mathcal{S}|_I$.*

**Proof.** Straightforward by the definition of the projection operation.   ◄

▶ **Lemma 34.** *Let $\mathcal{S}$ be an FBQS with quorum intersection. If $I_1$ and $I_2$ are sets such that $\mathcal{S}$ enjoys $I_1$- and $I_2$-availability, and $I_1$- and $I_2$-intersection, then the set $I = I_1 \cup I_2$ is such that $\mathcal{S}$ also enjoys $I$-availability and $I$-intersection.*

**Proof.** It is easy to check that $\mathcal{S}$ enjoys $I$-availability. We next prove that it also enjoys $I$-intersection. If one of $I_1$ and $I_2$ is empty, the property holds trivially. Hence, assume that $I_1$ and $I_2$ are non-empty, and thus are quorums in $\mathcal{S}$. Let $U_a$ and $U_b$ be any two quorums in $\mathcal{S}|_I$. Note that $U_a \cap I_1$ and $U_a \cap I_2$ cannot both be empty, for otherwise $(U_a \cap I_1) \cup (U_a \cap I_2) = U_a \cap I = U_a$ would be. Assume that $U_a \cap I_1 \neq \emptyset$. Then by Lemma 33, $U_a \cap I_1$ is a quorum in $(\mathcal{S}|_I)|_{I_1} = \mathcal{S}|_{I_1}$. Note that the set $U = I_1 \cap I_2$ is non-empty by quorum intersection in $\mathcal{S}$. But then, since $I_2$ is a quorum in $\mathcal{S}$, by Lemma 33, $U$ must be a quorum in $\mathcal{S}|_{I_1}$. Hence, both $U_a \cap I_1$ and $U$ are quorums in $\mathcal{S}|_{I_1}$, so that by $I_1$-intersection we have $\emptyset \neq (U_a \cap I_1) \cap U = U_a \cap I_1 \cap I_2$ and, thus, $U_a \cap I_2 \neq \emptyset$. Thus, in all cases we are guaranteed $U_a \cap I_2 \neq \emptyset$ and, hence, by Lemma 33, $U_a \cap I_2$ is a quorum in $\mathcal{S}|_{I_2}$. We can similarly show that $U_b \cap I_2$ is a quorum in $\mathcal{S}|_{I_2}$. But since $\mathcal{S}$ enjoys $I_2$-intersection, $(U_a \cap I_2) \cap (U_b \cap I_2) \neq \emptyset$, which is only possible if $U_a \cap U_b \neq \emptyset$. Therefore $\mathcal{S}$ enjoys $I$-intersection and the lemma holds. ◄

**Proof of Proposition 8.** By Lemma 34, the set of sets $I$ such that $\mathcal{S}$ enjoy $I$-correctness, $I$-availability and $I$-intersection is closed under union. Therefore, the biggest of such $I$'s is the union of all of them, which is unique. ◄

▶ **Lemma 35.** *Let $\mathcal{S}$ be an FBQS with quorum intersection. For a given set of faulty servers, let $\mathbf{V}_{\mathrm{int}}$ be the corresponding set of intact servers in $\mathcal{S}$. If $\mathbf{V}_{\mathrm{int}} \neq \emptyset$, then every quorum contains some intact server.*

**Proof.** Since $\mathbf{V}_{\mathrm{int}} \neq \emptyset$, $\mathbf{V}_{\mathrm{int}}$ is a quorum in $\mathcal{S}$. Since $\mathcal{S}$ has quorum intersection, $\mathbf{V}_{\mathrm{int}}$ intersects every other quorum in $\mathcal{S}$, which implies the required. ◄

**Proof of Proposition 9.** Consider two quorums $U_1$ and $U_2$ in $\mathcal{S}$. Let the set of intact servers in $\mathcal{S}$ be $\mathbf{V}_{\mathrm{int}} \neq \emptyset$. By Lemma 35 every quorum contains some intact server. Then both $U_1 \cap \mathbf{V}_{\mathrm{int}}$ and $U_2 \cap \mathbf{V}_{\mathrm{int}}$ are non-empty and, by Lemma 33, they are quorums in $\mathcal{S}|_{\mathbf{V}_{\mathrm{int}}}$. Since $\mathcal{S}|_{\mathbf{V}_{\mathrm{int}}}$ enjoys quorum intersection, $(U_1 \cap \mathbf{V}_{\mathrm{int}}) \cap (U_2 \cap \mathbf{V}_{\mathrm{int}}) = (U_1 \cap U_2) \cap \mathbf{V}_{\mathrm{int}} \neq \emptyset$. But then $U_1 \cap U_2$ must contain an intact server. ◄

**Proof of Theorem 10.** Let us fix a $B \in \mathcal{B}$. Since $B$ is maximal among the sets whose failure leaves the set of intact servers non-empty, the set $\mathbf{V} \setminus B$ is the set of intact servers assuming that the set of faulty servers is $B$. Hence, $\mathbf{V} \setminus B \neq \emptyset$, $\mathcal{S}|_{\mathbf{V} \setminus B}$ enjoys quorum intersection and $\mathbf{V} \setminus B$ is a quorum.

We now prove that $\forall U_1, U_2 \in \mathcal{Q}. U_1 \cap U_2 \not\subseteq B$, which establishes that $(\mathcal{Q}, \mathcal{B})$ satisfies D-consistency. By Lemma 33 we know that $U_1 \cap (\mathbf{V} \setminus B)$ and $U_2 \cap (\mathbf{V} \setminus B)$ are quorums in $\mathcal{S}|_{\mathbf{V} \setminus B}$. Since $\mathcal{S}|_{\mathbf{V} \setminus B}$ enjoys quorum intersection, $(U_1 \cap (\mathbf{V} \setminus B)) \cap (U_2 \cap (\mathbf{V} \setminus B)) = (U_1 \cap U_2) \setminus B \neq \emptyset$, and therefore $U_1 \cap U_2 \not\subseteq B$, as required.

We next prove that $\exists U \in \mathcal{Q}. B \cap U = \emptyset$, which establishes that $(\mathcal{Q}, \mathcal{B})$ satisfies D-availability. To this end, we let $U = \mathbf{V} \setminus B$. Then the required follows from the fact that $\mathbf{V} \setminus B$ is a quorum. ◄

**Proof of Corollary 11.** Let $\mathcal{S}$ be an FBQS with quorum intersection and let $(\mathcal{Q}, \mathcal{B})$ be the DQS it induces according to Theorem 10. Consider an execution of the protocol where at least one server in $\mathcal{S}$ is intact. Since $\mathcal{S}$ contains at least one intact server, by the definition of $\mathcal{B}$, some element of $\mathcal{B}$ contains all the faulty servers. Then the required follows from Theorem 3. ◄

## C  Proofs for §4

▶ **Lemma 36.** *Assume an FBQS $\mathcal{S}$ with quorum intersection and consider an execution of Stellar broadcast such that $\mathcal{S}$ contains some intact server. The first intact server that sends a* READY$(a)$ *message first needs to receive* ECHO$(a)$ *messages from every member of a quorum.*

**Proof.** The READY$(a)$ message may be sent at lines 11 or 14. The case of line 14 is impossible: since by Proposition 12 the set of befouled servers is not $v$-blocking for any intact server $v$, an intact server cannot be the first to send a READY$(a)$ message at this line. Hence, READY$(a)$ is sent at line 11. Then the required follows from the condition in the changed line 9. ◄

▶ **Lemma 37.** *Let $\mathcal{S}$ be an FBQS with quorum intersection. Consider an execution of the Stellar broadcast such that $\mathcal{S}$ contains some intact server. If some correct server delivers a value $a$, then some intact server has received ECHO($a$) messages from every member of a quorum.*

**Proof.** Assume that some correct server delivers value $a$. Then this server must have received a quorum of READY($a$) messages. By Lemma 35, at least one intact server must have sent a READY($a$) message. By Lemma 36, the first intact server to do so received a quorum of ECHO($a$) messages. ◀

**Proof of Lemma 16.** Assume that intact servers send messages READY($a$) and READY($a'$). By Lemma 36, some intact server $v$ has received ECHO($a$) from a quorum $U$, and some intact server $v'$ has received ECHO($a'$) from a quorum $U'$. By Proposition 9, $U \cap U'$ contains some intact server, so that this server has sent ECHO($a$) and ECHO($a'$). But due to the use of the guard variable echoed in lines 5 and 6, a server only echoes one value, and thus it cannot echo different values. Hence, $a = a'$. ◀

▶ **Lemma 38.** *Assume an FBQS $\mathcal{S}$ with quorum intersection, and let $B$ be a set of servers in $\mathbf{V}$. If $B$ is not $v$-blocking for any $v \in \mathbf{V} \setminus B$, then either $B = \mathbf{V}$ or $\mathbf{V} \setminus B$ is a quorum in $\mathcal{S}$.*

**Proof of Lemma 38.** Assume $B$ is not $v$-blocking for any $v \in \mathbf{V} \setminus B$. Then, for every server in $\mathbf{V} \setminus B$, there exists a slice $q \in \mathcal{S}(v)$ such that $q \cap B = \emptyset$, which implies that $q \subseteq \mathbf{V} \setminus B$. Then either $B = \mathbf{V}$ or $\mathbf{V} \setminus B$ is a quorum, as required. ◀

**Proof of Lemma 17.** Assume that $V^+$ is not $v$-blocking for any $v \in V^-$. By Lemma 38, either $V^- = \emptyset$ or $V^-$ is a quorum in $\mathcal{S}|_{\mathbf{V}_{\mathrm{int}}}$. In the former case we are done, while in the latter we get a contradiction as follows. By Lemma 33, $U \cap \mathbf{V}_{\mathrm{int}}$ is a quorum in $\mathcal{S}|_{\mathbf{V}_{\mathrm{int}}}$. Since $\mathcal{S}|_{\mathbf{V}_{\mathrm{int}}}$ enjoys quorum intersection, we have $(U \cap \mathbf{V}_{\mathrm{int}}) \cap V^- \neq \emptyset$. But this is impossible, since $U \cap \mathbf{V}_{\mathrm{int}} \subseteq V^+$ and $V^+ \cap V^- = \emptyset$. ◀

**Proof of Theorem 13.** *No duplication:* Straightforward by the use of the guard variable delivered in the changed line 15.

*Integrity:* Assume that some correct server delivers value $a$. By Lemma 37, some intact server received a quorum of ECHO($a$) messages. By Lemma 35, at least one intact server sent ECHO($a$). Since the sender is correct, this is only possible if the sender broadcast $a$.

*Consistency:* Assume that some correct server delivers value $a$ and some correct server delivers value $a'$. Then by the condition in the changed line 15, the servers received a quorum of READY($a$), respectively, READY($a'$) messages. By Lemma 35, either of these quorums contains some intact server. But then by Lemma 16 we have $a = a'$, as required.

*Validity for intact servers:* Assume the sender is correct and broadcasts $a$. Then every intact server will eventually send ECHO($a$). The set of intact servers $\mathbf{V}_{\mathrm{int}}$ is a quorum. Hence, every intact server will eventually send READY messages due to the condition in the changed line 9. By Lemma 16, these messages have to carry the value $a$. Then by the condition in the changed line 15, every intact server will eventually deliver a value. Due to the Consistency property, this value has to be $a$.

*Totality for intact servers:* Assume some correct server delivers value $a$. Then by the condition in the changed line 15 the server has received READY($a$) messages from every member in a quorum $U$. Since $U \cap \mathbf{V}_{\mathrm{int}}$ contains only intact servers, these servers send READY($a$) messages to every server. By the condition in the changed line 12, any correct server $v$ sends READY($a$) messages if it receives READY($a$) from every member in a $v$-blocking set. Hence, the READY($a$) messages from the servers in $U \cap \mathbf{V}_{\text{int}}$ may convince additional servers to send READY($a$) messages to every server. Let these additional servers send READY($a$) messages until a point is reached at which no further intact servers can send READY($a$) messages. At this point, let $V^+$ be the set of intact servers that sent READY($a$) messages (where $U \cap \mathbf{V}_{\text{int}} \subseteq V^+$), and let $V^- = \mathbf{V}_{\text{int}} \setminus V^+$. By Lemma 16 the servers in $V^-$ did not send any READY messages at all. The set $V^+$ cannot be $v$-blocking for any server $v$ in $V^-$, or else more intact servers could come to send READY($a$) messages. Then by Lemma 17 we have $V^- = \emptyset$, meaning that every intact server has sent READY($a$) messages. Since $\mathbf{V}_{\text{int}}$ is a quorum, all the intact servers will eventually deliver a value due to the condition in changed line 15. ◄

**Proof of Lemma 15.** By arguments similar to the ones in the proof of Theorem 13, the protocol enjoys the properties of No duplication, Consistency and Integrity. We prove that the protocol also enjoys the usual properties of Validity and Totality:

*Validity:* Assume the sender is correct and broadcasts $a$. By an argument similar to the one used for the property of Validity for intact servers in the proof of Theorem 13, every intact server will eventually deliver value $a$, which means that some correct server will eventually deliver value $a$. Now, by an argument similar to the one used for the property of Totality for intact servers in the proof of Theorem 13, every intact server will eventually send READY($a$). Then, by the condition in line 15, every correct server will eventually deliver a value. Due to the Consistency property, this value has to be $a$.

*Totality:* Assume some correct server delivers a value $a$. By an argument similar to the one used for the property of Totality for intact servers in the proof of Theorem 13, every intact server will eventually send READY($a$). Then, by the condition in line 15, every correct server will eventually deliver a value. ◄

▶ **Lemma 39.** *Let $\mathcal{S}$ be an FBQS with quorum intersection and let $(\mathcal{Q}, \mathcal{B})$ be the DQS that $\mathcal{S}$ induces. Let an execution of Bracha broadcast over $(\mathcal{Q}, \mathcal{B})$ with a history $H$ where some correct server delivers some value $a$. Then, there exists an execution of Stellar broadcast over $\mathcal{S}$ with the changed line 12 that also entails $H$.*

**Proof.** Since a correct server delivered $a$ in Bracha broadcast, by Lemma 30 some correct server $v$ received ECHO($a$) from a quorum $U$. This means that the first BCAST message received by every correct server in $U$ from the sender client is BCAST($a$), and thus $H$ contains the reception of such messages. By Totality, every correct server eventually delivers $a$ in Bracha broadcast, and therefore $H$ contains the triggering of such events. Now we construct an execution of Stellar broadcast that corresponds to $H$ as follows. We let every correct server in $U$ receive BCAST($a$) as the first BCAST message from the client sender. Therefore, every correct member of $U$ will send ECHO($a$) to every server in Stellar broadcast. Additionally, we let every faulty server in $U$ to also send ECHO($a$) messages to every server. Thus, by the conditions in line 9 of Figure 1, every correct server will eventually send READY($a$) to every server in Stellar broadcast. And by the condition in line 15, every correct server will eventually deliver $a$ in Stellar broadcast, which corresponds to $H$ and we are done. ◄

▶ **Lemma 40.** *Let $\mathcal{S}$ be an FBQS with quorum intersection and let $(\mathcal{Q}, \mathcal{B})$ be the DQS that $\mathcal{S}$ induces. Let an execution of Stellar broadcast over $\mathcal{S}$ with the changed line 12 that has a history $H$ where some correct server delivers some value $a$. Then, there exists an execution of Bracha broadcast over $(\mathcal{Q}, \mathcal{B})$ that also entails $H$.*

**Proof.** Since a correct server delivered $a$ in Stellar broadcast, by Lemma 37 some correct server $v$ received ECHO($a$) from a quorum $U$. This means that the first BCAST message received by every correct server in $U$ from the sender client is $\mathtt{BCAST}(a)$, and thus $H$ contains the reception of such messages. By Totality, every correct server eventually delivers $a$ in Stellar broadcast, and therefore $H$ contains the triggering of such events. Now we construct an execution of Bracha broadcast that corresponds to $H$ as follows. We let every correct server in $U$ receive $\mathtt{BCAST}(a)$ as the first $\mathtt{BCAST}$ message from the client sender. Therefore, every correct member of $U$ will send $\mathtt{ECHO}(a)$ to every server in Bracha broadcast. Additionally, we let every faulty server in $U$ to also send $\mathtt{ECHO}(a)$ messages to every server. Thus, by the condition in line 9 of Figure 1, every correct server will eventually send $\mathtt{READY}(a)$ to every server in Bracha broadcast. And by the condition in line 15, every correct server will eventually deliver $a$ in Bracha broadcast, which corresponds to $H$ and we are done.    ◀

▶ **Lemma 41.** *Let $\mathcal{S}$ be an FBQS with quorum intersection and let $(\mathcal{Q}, \mathcal{B})$ be the DQS that $\mathcal{S}$ induces. Let an execution of Bracha broadcast over $(\mathcal{Q}, \mathcal{B})$ with a history $H$ where no correct server ever delivers any value. Then, there exists an execution of Stellar broadcast over $\mathcal{S}$ with the changed line 12 that also entails $H$.*

**Proof.** We show that if no correct server ever delivers any value in the execution of Bracha broadcast with history $H$, then no correct server ever delivers any value in any execution of Stellar broadcast with history $H$ and where faulty servers never send any messages. Assume towards a contradiction that some correct server delivers $a$ in an execution of Stellar broadcast where faulty servers never send any messages. Since a correct server delivers $a$, by Lemma 37, some intact server receives $\mathtt{ECHO}(a)$ messages from every member of a quorum $U$ in Stellar broadcast. Since the faulty servers do not send any messages in Stellar broadcast, that means that every member of $U$ is correct, and thus $H$ contains the reception of $\mathtt{BCAST}(a)$ messages from the client sender by every member or $U$. But then, since the execution of Bracha broadcast has history $H$, every member of $U$ will send $\mathtt{ECHO}(a)$ in Bracha broadcast, and by the conditions in lines 9 and 15, every correct server will eventually send $\mathtt{READY}(a)$ to every server, and will eventually deliver $a$, which contradicts our assumptions. Therefore, no correct server will ever deliver $a$ in any execution of Stellar broadcast that has history $H$ and where faulty servers never send any messages, and the lemma holds.    ◀

▶ **Lemma 42.** *Let $\mathcal{S}$ be an FBQS with quorum intersection and let $(\mathcal{Q}, \mathcal{B})$ be the DQS that $\mathcal{S}$ induces. Let an execution of Stellar broadcast over $\mathcal{S}$ with the changed line 12 that has a history $H$ where no correct server ever delivers any value. Then, there exists an execution of Bracha broadcast over $(\mathcal{Q}, \mathcal{B})$ that also entails $H$.*

**Proof.** We show that if no correct server ever delivers any value in the execution of Stellar broadcast with history $H$, then no correct server ever delivers any value in any execution of Bracha broadcast with history $H$ and where faulty servers never send any messages. Assume towards a contradiction that some correct server delivers $a$ in an execution of Bracha broadcast where faulty servers never send any messages. Since a correct server delivers $a$, by Lemma 30, some intact server receives $\mathtt{ECHO}(a)$ messages from every member of a quorum $U$ in Bracha broadcast. Since the faulty servers do not send any messages in Bracha broadcast, that means that every member of $U$ is correct, and thus $H$ contains the reception of $\mathtt{BCAST}(a)$ messages from the client sender by every member or $U$. But then, since the execution of Stellar broadcast has history $H$, every member of $U$ will send $\mathtt{ECHO}(a)$ in Stellar broadcast, and by the conditions in lines 9 and 15, every correct server will eventually send $\mathtt{READY}(a)$ to every server, and will eventually deliver $a$, which contradicts our assumptions. Therefore, no correct server will ever deliver $a$ in any execution of Bracha broadcast with history $H$ and where faulty servers never send any messages, and the lemma holds.    ◀

**Proof of Theorem 18.** The theorem is a straightforward consequence of Lemmas 39– 42.
◀

## D  Proofs for §5

▶ **Lemma 43.** *Let* $\{\mathcal{S}_v\}_{v \in \mathbf{V}_{\mathrm{ok}}}$ *be a subjective FBQS with quorum intersection. Assume that* $I_1$ *and* $I_2$ *are sets such that for every* $v \in \mathbf{V}_{\mathrm{ok}}$, $\mathcal{S}$ *enjoys* $I_1$- *and* $I_2$-*availability, and* $I_1$- *and* $I_2$-*intersection. Then the set* $I = I_1 \cup I_2$ *is such that for every* $v \in \mathbf{V}_{\mathrm{ok}}$, $\mathcal{S}_v$ *also enjoys* $I$-*availability and* $I$-*intersection.*

**Proof.** The required follows from Lemma 34. ◀

**Proof of Proposition 20.** Let $I$ be a set such that for every $v \in \mathbf{V}_{\mathrm{ok}}$, $\mathcal{S}_v$ enjoys $I$-correctness, that is, $I$ contains only correct servers. Then for every $v_1$ and $v_2$ in $\mathbf{V}_{\mathrm{ok}}$ we have $\mathcal{S}_{v_1}|_I = \mathcal{S}_{v_2}|_I$. If $I$ is a quorum in one of the views $\mathcal{S}_{v_1}$ and $\mathcal{S}_{v_2}$, then $I$ is also a quorum in the other view. Similarly, if one of $\mathcal{S}_{v_1}|_I$ and $\mathcal{S}_{v_2}|_I$ enjoys quorum intersection, then so does the other. Thus, if some view $\mathcal{S}_v$ with $v \in \mathbf{V}_{\mathrm{ok}}$ enjoys $I$-correctness, $I$-availability and $I$-intersection, then any other view $\mathcal{S}_{v'}$ with $v' \in \mathbf{V}_{\mathrm{ok}}$ will also enjoy those properties. By Lemma 43, the set of intact nodes is the union of all sets $I$ such that for every $v \in \mathbf{V}_{\mathrm{ok}}$, $\mathcal{S}_v$ enjoys $I$-correctness, $I$-availability, and $I$-intersection. The above-established property implies that this coincides with the set of intact servers in each $\mathcal{S}_v$ with $v \in \mathbf{V}_{\mathrm{ok}}$. ◀

▶ **Lemma 44.** *Let* $\{\mathcal{S}_v\}_{v \in \mathbf{V}_{\mathrm{ok}}}$ *be a subjective FBQS with quorum intersection and at least one intact server, and let* $\{\mathcal{Q}_v\}_{v \in \mathbf{V}_{\mathrm{ok}}}$ *be its induced subjective quorum system. Then every quorum in* $\bigcup_{v \in \mathbf{V}_{\mathrm{ok}}} \mathcal{Q}_v$ *contains some intact server.*

**Proof.** Since the set of intact servers $\mathbf{V}_{\mathrm{int}}$ is non-empty, $\mathbf{V}_{\mathrm{int}}$ is a quorum in $\mathcal{S}_v$ for each $v \in \mathbf{V}_{\mathrm{ok}}$. Since each $\mathcal{S}_v$ has quorum intersection, $\mathbf{V}_{\mathrm{int}}$ intersects every quorum in $\bigcup_{v \in \mathbf{V}_{\mathrm{ok}}} \mathcal{Q}_v$, as required. ◀

**Proof of Proposition 21.** Consider two quorums $U_1$ and $U_2$ in $\bigcup_{v \in \mathbf{V}_{\mathrm{ok}}} \mathcal{S}_v$. Let the set of intact servers in $\{\mathcal{S}_v\}_{v \in \mathbf{V}_{\mathrm{ok}}}$ be $\mathbf{V}_{\mathrm{int}} \neq \emptyset$. By Proposition 20, $\mathbf{V}_{\mathrm{int}}$ coincides with the set of intact servers in each of the views $\mathcal{S}_v$ with $v \in \mathbf{V}_{\mathrm{ok}}$. By Lemma 44 every quorum contains some intact server. Then both $U_1 \cap \mathbf{V}_{\mathrm{int}}$ and $U_2 \cap \mathbf{V}_{\mathrm{int}}$ are non-empty. Let $v_1$ and $v_2$ be such that $U_1$ is a quorum in $\mathcal{S}_{v_1}$ and $U_2$ a quorum in $\mathcal{S}_{v_2}$. By Lemma 33, $U_1 \cap \mathbf{V}_{\mathrm{int}}$ is a quorum in $\mathcal{S}_{v_1}|_{\mathbf{V}_{\mathrm{int}}}$ and $U_2 \cap \mathbf{V}_{\mathrm{int}}$ is a quorum in $\mathcal{S}_{v_2}|_{\mathbf{V}_{\mathrm{int}}}$. Since the views in a subjective FBQS agree on the correct servers, then $\mathcal{S}_{v_1}|_{\mathbf{V}_{\mathrm{int}}} = \mathcal{S}_{v_2}|_{\mathbf{V}_{\mathrm{int}}}$. Since $\mathcal{S}_{v_1}|_{\mathbf{V}_{\mathrm{int}}}$ enjoys quorum intersection, $(U_1 \cap \mathbf{V}_{\mathrm{int}}) \cap (U_2 \cap \mathbf{V}_{\mathrm{int}}) = (U_1 \cap U_2) \cap \mathbf{V}_{\mathrm{int}} \neq \emptyset$. But then $U_1 \cap U_2$ must contain an intact server. ◀

▶ **Proposition 45.** *Assume a subjective FBQS* $\{\mathcal{S}_v\}_{v \in \mathbf{V}_{\mathrm{ok}}}$ *with quorum intersection. Its set of befouled servers* $V_{\mathrm{bef}}$ *is not* $v$-*blocking for any intact* $v$.

**Proof.** Consider an intact server $v$. The set of all intact servers $\mathbf{V}_{\mathrm{int}} = \mathbf{V} \setminus \mathbf{V}_{\mathrm{bef}}$ is a quorum in $\mathcal{S}_v$. Then there exists a slice $q \in \mathcal{S}_v(v)$ such that $q \subseteq \mathbf{V}_{\mathrm{int}}$. Hence, $\mathbf{V}_{\mathrm{bef}}$ cannot overlap all of $v$'s slices known by $v$ itself, as required. ◀

▶ **Lemma 46.** *Assume a subjective FBQS* $\{\mathcal{S}_v\}_{v \in \mathbf{V}_{\mathrm{ok}}}$ *with quorum intersection and consider an execution of Stellar broadcast such that* $\{\mathcal{S}_v\}_{v \in \mathbf{V}_{\mathrm{ok}}}$ *contains some intact server. The first intact server* $v$ *that sends a* `READY`(a) *message first needs to receive* `ECHO`(a) *messages from every member of a quorum in* $\mathcal{S}_v$.

**Proof.** The `READY`($a$) message may be sent at lines 11 or 14. The case of line 14 is impossible: since by Proposition 45 the set of befouled servers is not $v$-blocking for any intact server $v$, an intact server cannot be the first to send a `READY`($a$) message at this line. Hence, `READY`($a$) is sent at line 11. Then the required follows from the condition in the changed line 9. ◀

▶ **Lemma 47.** *Let $\{\mathcal{S}_v\}_{v \in \mathbf{V}_{ok}}$ be a subjective FBQS with quorum intersection where at least one server is intact. If some correct server in an execution of the Stellar broadcast delivers a value $a$, then some intact server $v$ has received `ECHO`($a$) messages from every member of a quorum known by $v$.*

**Proof.** Assume that some correct server delivers a value $a$. Then this server must have received `READY`($a$) messages from a quorum it knows. By Lemma 44, at least one intact server must have sent a `READY`($a$) message. By Lemma 46, the first intact server $v$ to do so received `ECHO`($a$) messages from a quorum known by $v$. ◀

**Proof of Lemma 23.** Assume that intact servers send messages `READY`($a$) and `READY`($a'$). By Lemma 46, some intact server $v$ has received `ECHO`($a$) from a quorum $U$ known by $v$, and some intact server $v'$ has received `ECHO`($a'$) from a quorum $U'$ known by $v'$. By Proposition 21, $U \cap U'$ contains some intact server, so that this server has sent `READY`($a$) and `READY`($a'$). But due to the use of the guard variable `echoed` in lines 5 and 6, a server only echoes one value, and thus it cannot echo different values. Hence, $a = a'$. ◀

**Proof of Lemma 24.** Since every intact server is correct, all the views $\mathcal{S}_v|_{\mathbf{V}_{int}}$ with $v \in \mathbf{V}_{ok}$ coincide with each other. Since $V^+$ and $V^-$ only contain intact servers, they both lie within the subsystem $\mathcal{S}_v|_{\mathbf{V}_{int}}$. Thus, for every $v \in V^+$, $V^-$ is $v$-blocking in $\{\mathcal{S}_v\}_{v \in \mathbf{V}_{ok}}$—in the sense of §5—iff $V^-$ is $v$-blocking in any $\mathcal{S}_{v'}|_{\mathbf{V}_{int}}$ with $v' \in \mathbf{V}_{ok}$—in the sense of §4. Assume that $V^+$ is not $v$-blocking for any $v \in V^-$ in $\{\mathcal{S}_v\}_{v \in \mathbf{V}_{ok}}$. Then $V^+$ is not $v$-blocking for any $v \in V^-$ in some $\mathcal{S}_{v'}|_{\mathbf{V}_{int}}$ with $v' \in \mathbf{V}_{ok}$. By Lemma 38, either $V^- = \emptyset$ or $V^-$ is a quorum in $\mathcal{S}_{v'}|_{\mathbf{V}_{int}}$. In the former case we are done, while in the latter we get a contradiction as follows. By Lemma 33, $U \cap \mathbf{V}_{int}$ is a quorum in $\mathcal{S}_v|_{\mathbf{V}_{int}} = \mathcal{S}_{v'}|_{\mathbf{V}_{int}}$. Since $\mathcal{S}_{v'}|_{\mathbf{V}_{int}}$ enjoys quorum intersection, we have $U \cap \mathbf{V}_{int} \cap V^- \neq \emptyset$. But this is impossible, since $U \cap \mathbf{V}_{int} \subseteq V^+$ and $V^+ \cap V^- = \emptyset$. ◀

**Proof of Theorem 22.** *No duplication:* Straightforward by the use of the guard variable `delivered` in the changed line 15.

*Integrity:* Assume that some correct server delivers value $a$. By Lemma 47, some intact server received `ECHO`($a$) messages form a quorum it knows. By Lemma 44, at least one intact server sent `ECHO`($a$). Since the sender is correct, this is only possible if the sender broadcast $a$.

*Consistency:* Assume that some correct server delivers value $a$ and some correct server delivers value $a'$. Then by the condition in the changed line 15, the servers received `READY`($a$) and `READY`($a'$) from quorums known by each of the servers, respectively. By Lemma 44, either of these quorums contains some intact server. But then by Lemma 23 we have $a = a'$, as required.

*Validity for intact servers:* Assume the sender is correct and broadcasts $a$. Then every intact server will eventually send `ECHO`($a$). The set of intact servers $\mathbf{V}_{int}$ is a quorum in $\mathcal{S}_v$ for every $v \in \mathbf{V}_{ok}$. Hence, every intact server will eventually send `READY` messages due to the condition in the changed line 9. By Lemma 23, these messages have to carry the value $a$. Then by the condition in the changed line 15, every intact server will eventually deliver a value. Due to the Consistency property, this value has to be $a$.

*Totality for intact servers:* Assume some correct server delivers value $a$. Then by the condition in the changed line 15 the server has received $\texttt{READY}(a)$ messages from every member in a quorum $U$ it knows. Since $U \cap \mathbf{V}_{\text{int}}$ contains only intact servers, these servers send $\texttt{READY}(a)$ messages to every server. By the condition in the changed line 12, any correct server $v$ sends $\texttt{READY}(a)$ messages if it receives $\texttt{READY}(a)$ from every member in a $v$-blocking set. Hence, the $\texttt{READY}(a)$ messages from the servers in $U \cap \mathbf{V}_{\text{int}}$ may convince additional servers to send $\texttt{READY}(a)$ messages to every server. Let these additional servers send $\texttt{READY}(a)$ messages until a point is reached at which no further intact servers can send $\texttt{READY}(a)$ messages. At this point, let $V^+$ be the set of intact servers that sent $\texttt{READY}(a)$ messages (where $U \cap \mathbf{V}_{\text{int}} \subseteq V^+$), and let $V^- = \mathbf{V}_{\text{int}} \setminus V^+$. By Lemma 23 the servers in $V^-$ did not send any $\texttt{READY}$ messages at all. The set $V^+$ cannot be $v$-blocking for any server $v$ in $V^-$, or else more intact servers could come to send $\texttt{READY}(a)$ messages. Then by Lemma 24 we have $V^- = \emptyset$, meaning that every intact server has sent $\texttt{READY}(a)$ messages. Since $\mathbf{V}_{\text{int}}$ is a quorum in $\mathcal{S}_v$ for every $v \in \mathbf{V}_{\text{ok}}$, all the intact servers will eventually deliver a value due to the condition in changed line 15. ◀

## **E** Proofs for §6

**Proof of Theorem 25.** Each pair $(\mathcal{Q}_v, \mathcal{B}_v)$ with $v \in \mathbf{V}_{\text{ok}}$ is a DQS, and thus it satisfies D-availability. For each $v \in \mathbf{V}_{\text{ok}}$, the fail-prone system $\mathcal{B}_v$ contains the maximal sets such that their failure leaves the set of intact servers in $\mathcal{Q}_v$ non-empty. Since $\{\mathcal{S}_v\}_{v \in \mathbf{V}_{\text{ok}}}$ contains some intact server, then the set $\mathbf{V}_{\text{bef}}$ of befouled servers is the only element that contains all faulty servers in each of the fail-prone systems $\mathcal{B}_v$ with $v \in \mathbf{V}_{\text{ok}}$. This entails that $(\mathcal{Q}_v, \mathcal{B}_v)$ enjoys SD-safety. By the construction of $\{\mathcal{Q}_v, \mathcal{B}\}_{v \in \mathbf{V}_{\text{ok}}}$, for every $v \in \mathbf{V}_{\text{ok}}$ the only element $B \in \mathcal{B}_v$ that contains all faulty servers, only contains befouled servers. SD-consistency holds because by Proposition 21 the intersection of any two quorums, coming from the same or different views, contains some intact server. ◀

▶ **Lemma 48.** *Consider a subjective DQS* $(\{\mathcal{Q}_v\}_{v \in \mathbf{V}_{\text{ok}}}, \{\mathcal{B}\}_{v \in \mathbf{V}_{\text{ok}}})$. *Then for every* $v \in \mathbf{V}_{\text{ok}}$, *every quorum in* $\mathcal{Q}_v$ *contains some correct server.*

**Proof.** Let us fix a $v \in \mathbf{V}_{\text{ok}}$. By SD-safety some element $B \in \mathcal{B}_v$ contains all faulty servers. Then by D-availability in $(\mathcal{Q}_v, \mathcal{B}_v)$, there is a quorum $U$ in $\mathcal{Q}_v$ that contains only correct servers. Consider an arbitrary quorum $U'$ in $\mathcal{Q}_v$. Since $\mathcal{Q}_v$ is a quorum system, the quorums $U$ and $U'$ intersect. Then, since $U$ contains only correct servers, $U'$ contains some correct server. ◀

▶ **Lemma 49.** *Assume a subjective DQS* $(\{\mathcal{Q}_v\}_{v \in \mathbf{V}_{\text{ok}}}, \{\mathcal{B}\}_{v \in \mathbf{V}_{\text{ok}}})$ *and consider an execution of the protocol in Figure 1. The first correct server* $v$ *that sends a* $\texttt{READY}(a)$ *message first needs to receive* $\texttt{ECHO}(a)$ *messages from every member of a quorum known by* $v$.

**Proof.** The $\texttt{READY}(a)$ message may be sent at lines 11 or 14. The case of line 14 is impossible: since some element of $\mathcal{B}_v$ contains all the faulty servers, a correct server cannot be the first to send a $\texttt{READY}(a)$ message at this line. Hence, $\texttt{READY}(a)$ is sent at line 11. Then the required follows from the condition in line 9. ◀

▶ **Lemma 50.** *Assume a subjective DQS* $(\{\mathcal{Q}_v\}_{v \in \mathbf{V}_{\text{ok}}}, \{\mathcal{B}\}_{v \in \mathbf{V}_{\text{ok}}})$ *and consider an execution of the protocol in Figure 1. If some correct server delivers a value* $a$, *then some correct server* $v$ *received* $\texttt{ECHO}(a)$ *messages from every member of a quorum known by* $v$.

**Proof.** Assume that some correct server delivers value $a$. Then this server must have received `READY`($a$) messages from a quorum known by the server. By Lemma 48, at least one correct server must have sent a `READY`($a$) message. By Lemma 49, the first server $v$ to do so received `ECHO`($a$) messages from a quorum known by $v$. ◁

▶ **Lemma 51.** *Assume a subjective DQS* $(\{\mathcal{Q}_v\}_{v \in \mathbf{V}_{ok}}, \{\mathcal{B}_{v \in \mathbf{V}_{ok}}\})$ *and consider an execution of the protocol in Figure 1. If correct servers send messages* `READY`($a$) *and* `READY`($a'$)*, then* $a = a'$.

**Proof.** Assume that correct servers send messages `READY`($a$) and `READY`($a'$). Consider the first correct servers that send the respective messages, $v$ and $v'$. By Lemma 49, $v$ has received `ECHO`($a$) messages from a quorum $U \in \mathcal{Q}_v$, and $v'$ has received `ECHO`($a'$) messages from a quorum $U' \in \mathcal{Q}_{v'}$. By SD-safety, some element $B \in \mathcal{B}_v$ contains all faulty servers. By SD-consistency, $U \cap U' \not\subseteq B$, and thus $U \cap U'$ contains some correct server, so that this server has sent `ECHO`($a$) and `ECHO`($a'$). But due to the use of the guard variable echoed in lines 5 and 7, a server only echoes one value, and thus it cannot echo different values. Hence, $a = a'$, as required. ◁

▶ **Lemma 52.** *Consider a subjective DQS* $(\{\mathcal{Q}_v\}_{v \in \mathbf{V}_{ok}}, \{\mathcal{B}\}_{v \in \mathbf{V}_{ok}})$*. Let* $v \in \mathbf{V}_{ok}$ *and let* $U$ *be a quorum in* $\mathcal{Q}_v$*. Let* $U^+$ *be the set of correct servers in* $U$*. Then for every* $v' \in \mathbf{V}_{ok}$ *and every* $B' \in \mathcal{B}_{v'}$ *we have* $U^+ \not\subseteq B'$.

**Proof.** Assume towards a contradiction that for some $v' \in \mathbf{V}_{ok}$ and $B' \in \mathcal{B}_{v'}$ we have $U^+ \subseteq B'$. By D-availability in $(\mathcal{Q}_{v'}, \mathcal{B}_{v'})$ we know that $\mathcal{Q}_{v'}$ contains a quorum $U' \subseteq \mathbf{V} \setminus U^+$. By SD-safety, some element $B \in \mathcal{B}_v$ contains all faulty servers. Since $U^+$ only contains correct servers, $(U \cap U') \setminus B \subseteq U^+ \cap (U' \setminus B) \subseteq U^+ \cap U'$. But $U^+ \cap U' = \emptyset$, which implies that $U \cap U' \subseteq B$, thus contradicting SD-consistency. ◁

**Proof of Theorem 26.** *No duplication:* Straightforward by the use of the guard variable delivered in line 15.

*Integrity:* Assume that some correct server delivers value $a$. By Lemma 50, some correct server received `ECHO`($a$) messages from a quorum known by that correct server. Then by Lemma 48, at least one correct server sent `ECHO`($a$). Since the sender is correct, this is only possible if the sender broadcast $a$.

*Consistency:* Assume that some correct server delivers value $a$, and some correct server delivers value $a'$. Then by the condition in line 15, the servers received `READY`($a$) and `READY`($a'$) messages form quorums known by each of the servers respectively. By Lemma 48, either of these quorums contains some correct server. But then by Lemma 51 we have $a = a'$, as required.

*Validity:* Assume the sender is correct and broadcasts $a$. Then every correct server will eventually send `ECHO`($a$). For every $v \in \mathbf{V}_{ok}$ there exists a quorum $U_v \in \mathcal{Q}_v$ consisting only of correct servers, due to SD-safety and D-availability in $(\mathcal{Q}_v, \mathcal{B}_v)$. Hence, all members of each $U_v$ will send `ECHO`($a$). Hence, every correct server will eventually send `READY` messages due to the condition in line 9. By Lemma 51, all these messages must carry the value $a$. Then, since there exists a quorum consisting only of correct servers in each $\mathcal{Q}_v$, by the condition in line 15, every correct server will eventually deliver a value. Due to the Consistency property, this value has to be $a$.

*Totality:* Assume some correct server delivers value $a$. Then by the condition in line 15 the server has received `READY`($a$) messages from every member in a quorum $U$ it knows. Consider the set $U^+$ that contains the correct servers in $U$. By Lemma 48, $U^+ \neq \emptyset$.

Every server in $U^+$ sends `READY`($a$) to every server. By Lemma 52, for every $v \in \mathbf{V}_{\mathrm{ok}}$ and every $B' \in \mathcal{B}_v$ we have $U^+ \nsubseteq B'$. Therefore, every correct server that has not sent `READY` messages already will eventually do so due to the condition in line 12. Due to Lemma 51, all these messages must carry the value $a$. Thus, every correct server will eventually send `READY`($a$). For every $v \in \mathbf{V}_{\mathrm{ok}}$ there exists a quorum $U_v \in \mathcal{Q}_v$ consisting only of correct servers, due to SD-safety and D-availability in $(\mathcal{Q}_v, \mathcal{B}_v)$. Therefore, all correct servers will eventually deliver a value due to the condition in line 15.   ◄

**Proof of Corollary 27.** By Theorems 25 and 26.   ◄